\documentclass[a4paper, 11pt, english]{article}

\usepackage[utf8]{inputenc}
\usepackage[T1]{fontenc}
\usepackage{lmodern}

\usepackage{amsthm, amssymb, amsmath}

\usepackage[top=2.5cm, bottom=2.5cm, left=2.5cm, right=2.5cm]{geometry}
\usepackage{graphicx}
\usepackage{microtype}
\usepackage{enumerate}
\usepackage[font=small,labelfont=bf]{caption}
\usepackage{cite} 
\usepackage{tikz}
\usepackage{xcolor}
\usepackage{siunitx} 
\usepackage{multirow} 
\usepackage[titletoc,title]{appendix}

\theoremstyle{remark}

\definecolor{rouge}{rgb}{0.85,0.11,0.07}
\definecolor{vert}{rgb}{0.25,0.75,0.15}
\definecolor{bleu}{rgb}{0.0,0.36,0.80}
\definecolor{orange}{rgb}{0.94,0.43,0.02}
\definecolor{bleugris}{rgb}{0.03,0.35,0.58}
\definecolor{bleuclair}{rgb}{0.23,0.46,0.80}
\definecolor{rose}{rgb}{0.8275,0.2118,0.5098}

\usepackage[colorlinks=true]{hyperref}
\hypersetup{linkcolor=black, citecolor=bleugris, urlcolor = black}

\usepackage[numbered]{bookmark}

{ \begin{list}%
	{\raisebox{-0ex}{•}}%
	{\setlength{\labelsep}{3pt}%
	 \setlength{\leftmargin}{18pt}%
	 \setlength{\itemsep}{\parsep}}}%
{ \end{list} }

\newcommand*{\R}{\ensuremath \mathbb{R}} 

\newcommand*{\dr}{\ensuremath \partial} 
\newcommand*{\grad}{\ensuremath {\nabla}} 

\newcommand*{\Bb}{\ensuremath \mathcal{B}} 
\newcommand*{\Ee}{\ensuremath \mathcal{E}} 
\newcommand*{\Ff}{\ensuremath \mathcal{F}} 
\newcommand*{\Gg}{\ensuremath \mathcal{G}} 
\newcommand*{\Ll}{\ensuremath \mathcal{L}} 
\newcommand*{\Mm}{\ensuremath \mathcal{M}} 
\newcommand*{\Pp}{\ensuremath \mathcal{P}} 
\newcommand*{\Ss}{\ensuremath \mathcal{S}} 

\newcommand*{\intd}[1]{\ensuremath \operatorname{d}\!{#1}} 

\newcommand*{\kon}{\ensuremath k_{\text{\normalfont on}}}
\newcommand*{\koff}{\ensuremath k_{\text{\normalfont off}}}

\newcommand*{\koni}{\ensuremath k_{\text{\normalfont on},i}}
\newcommand*{\koffi}{\ensuremath k_{\text{\normalfont off},i}}
\newcommand*{\Koni}{\ensuremath K_{\text{\normalfont on},i}}
\newcommand*{\Koffi}{\ensuremath K_{\text{\normalfont off},i}}

\newcommand*{\Bet}{\ensuremath \operatorname{Beta}}
\newcommand*{\Beta}{\ensuremath \operatorname{B}}

\newcommand*{\xx}{\ensuremath \mathbf{x}}
\newcommand*{\yy}{\ensuremath \mathbf{y}}

\renewcommand{\baselinestretch}{1.1}

\title{\vspace{-15mm} \textbf{Inferring gene regulatory networks from single-cell~data: a mechanistic approach}}
\date{}
\author{Ulysse Herbach\textsuperscript{1,2,3}\footnote{\href{mailto:ulysse.herbach@ens-lyon.fr}{ulysse.herbach@ens-lyon.fr}}, Arnaud Bonnaffoux\textsuperscript{1,2,4}, Thibault Espinasse\textsuperscript{3}, Olivier Gandrillon\textsuperscript{1,2}}

\begin{document}

\maketitle

\vspace{-7mm}

\begin{center}
{\small
\textsuperscript{1} Univ Lyon, ENS de Lyon, Univ Claude Bernard, CNRS UMR 5239, INSERM U1210, Laboratory of Biology and Modelling of the Cell, 46 all\'{e}e d'Italie Site Jacques Monod, F-69007 Lyon, France

\textsuperscript{2} Inria Team Dracula, Inria Center Grenoble Rhône-Alpes, France

\textsuperscript{3} Univ Lyon, Universit\'{e} Claude Bernard Lyon 1, CNRS UMR 5208, Institut Camille Jordan, 43 blvd. du 11 novembre 1918, F-69622 Villeurbanne cedex, France

\textsuperscript{4} The CoSMo company, 5 passage du Vercors, 69007 Lyon, France}
\end{center}

{\abstract

The recent development of single-cell transcriptomics has enabled gene expression to be measured in individual cells instead of being population-averaged. Despite this considerable precision improvement, inferring regulatory networks remains challenging because stochasticity now proves to play a fundamental role in gene expression. In particular, mRNA synthesis is now acknowledged to occur in a highly bursty manner.
We propose to view the inference problem as a fitting procedure for a mechanistic gene network model that is inherently stochastic and takes not only protein, but also mRNA levels into account. We first explain how to build and simulate this network model based upon the coupling of genes that are described as piecewise-deterministic Markov processes. Our model is modular and can be used to implement various biochemical hypotheses including causal interactions between genes. However, a naive fitting procedure would be intractable. By performing a relevant approximation of the stationary distribution, we derive a tractable procedure that corresponds to a statistical hidden Markov model with interpretable parameters. This approximation turns out to be extremely close to the theoretical distribution in the case of a simple toggle-switch, and we show that it can indeed fit real single-cell data. As a first step toward inference, our approach was applied to a number of simple two-gene networks simulated in silico from the mechanistic model and satisfactorily recovered the original networks.
Our results demonstrate that functional interactions between genes can be inferred from the distribution of a mechanistic, dynamical stochastic model that is able to describe gene expression in individual cells. This approach seems promising in relation to the current explosion of single-cell expression data.

}

\section{Introduction}

Inferring regulatory networks from gene expression data is a longstanding question in systems biology~\cite{Hecker2009}, with an active community developing many possible solutions. So far, almost all studies have been based on population-averaged data, which historically used to be the only possible way to observe gene expression. Technologies now allow us to measure mRNA levels in individual cells~\cite{Kanter2015,Tang2009,Wagner2016}, a revolution in terms of precision. However, the network reconstruction task paradoxically remains more challenging than ever.

\bigskip

The main reason is that the variability in gene expression unexpectedly stands at a large distance from a trivial, limited perturbation around the population mean. It is now clear indeed that this variability can have functional significance~\cite{Huang2009,Eldar2010,Dueck2015} and should therefore not be ignored when dealing with gene network inference. In particular, as the mean is not sufficient to account for a population of cells, a deterministic model -- e.g. ordinary differential equation (ODE) systems, often used in inference~\cite{Mizeranschi2015,Matsumoto2017} -- is unlikely to faithfully inform about an underlying gene regulatory network. Whether such a deterministic approach could still be a valid approximation or not is a difficult question that may require some biological insight into the system under consideration~\cite{Symmons2016}.
Another key aspect when considering individual cells is that they generally have to be killed for measurements: from a statistical point of view, temporal single-cell data therefore should not be seen as a set of time series, but rather \emph{snapshots}, i.e. independent samples from a time series of distributions.

\bigskip

On the other hand, single-cell data give the opportunity of moving one step further toward a more accurate physical description of gene expression. Molecular processes of gene expression are overall now well understood, in particular transcription, but precisely how stochasticity emerges is still somewhat of a conundrum. Harnessing variability in single-cell data is expected to allow for the identification of critical parameters and also to provide hints about the basic molecular processes involved~\cite{Munsky2009,Zimmer2015}.
Moreover, the variability arising from perturbations in cell populations is often crucial for network reconstruction to succeed~\cite{Cai2013,Djordjevic2014} as the deterministic inference problem suffers from intrinsic limitations~\cite{Angulo2017}. From this point of view, the same information is expected to be contained in the variability between cells in single-cell data.
Some of the few existing single-cell inference methods follow this path, for example using asynchronous Boolean network models \cite{Moignard2015} or generating pseudo time series~\cite{Ocone2015,Matsumoto2017}.
In this article, we use a mechanistic approach in the sense that every part of our model has an explicit physical interpretation. Importantly, mRNA observations are not used as a proxy for proteins since both are explicitly modeled.

\bigskip

Besides, mechanistic models that are accurate enough to describe gene expression at the single-cell level usually do not consider interactions between genes. For example, the so-called “two-state” (aka random telegraph) model has been successfully used with single-cell RNA-seq data~\cite{Kim2013}, but the joint distribution of a set of genes contains much more information than the marginal kinetics of individual genes: our aim is to exploit this information while keeping the mechanistic point of view.

\bigskip

{Namely, we propose to view the inference as a fitting procedure for a mechanistic gene network model. Whereas the goal here is not to achieve global predictability performances (e.g. as in~\cite{Marbach2012}), our framework makes it possible to explicitly implement many biological hypotheses, and to test them by going back and forth between simulations and experiments. The main point of this article is to show that a tractable statistical model for network inference from single-cell data can be derived through successive relevant approximations. Finally, we demonstrate that our approach is capable of extracting enough information out of in silico-simulated noisy single-cell data to correctly infer the structures of various two-gene networks.}

\section{Methods}

In this part, we aim at deriving a tractable statistical model from a mechanistic one. We will use the two-state model for gene expression to build a “network of two-state models” by making the promoter switching rates depend on protein levels.
Then, successive relevant simplifications will lead to an explicit approximation of a statistical likelihood.

\subsection{A simple mechanistic model for gene regulatory networks}

\subsubsection{Basic block: stochastic expression of a single gene}

\begin{figure}[ht]
\begin{center}
\includegraphics[width=0.6\textwidth]{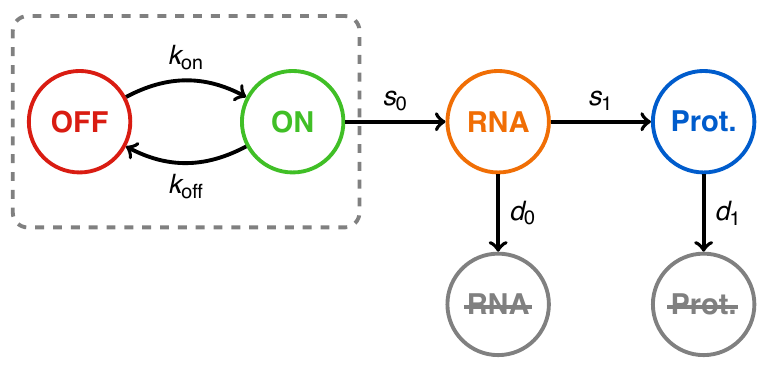}
\caption{Scheme of the two-state model of gene expression, basic block of our network model.}
\label{scheme_two_state}
\end{center}
\end{figure}

Our starting point is the well-known two-state model of gene expression~\cite{Raser2004,Becskei2005,Raj2006,Suter2011}, a refinement of the model introduced by~\cite{Ko1991} from pioneering single-cell experiments~\cite{Ko1990}.
In this model, a gene is described by its promoter which can be either active (on) or inactive (off) -- possibly representing a transcription complex being “bound” or “unbound” but it may be more complicated~\cite{Larson2011} -- with mRNA being transcribed only during the active periods. Translation is added in a standard way, each mRNA molecule producing proteins at a constant rate. The resulting model (Fig.~\ref{scheme_two_state}) can be entirely defined by the set of chemical reactions detailed in Table~\ref{parameters_single}, where chemical species $G$, $G^*$, $M$ and $P$ respectively denote the inactive promoter, the active promoter, the amount of mRNA and proteins.
The mathematical framework generally assumes stochastic mass-action kinetics~\cite{Gillespie1977} for all reactions, since they typically involve few molecules compared to Avogadro's number.
In this fully discrete setting, one can use the master equation to compute stationary distributions: for mRNA the exact distribution is a Beta-Poisson mixture~\cite{Dattani2017}, and an approximation is available for proteins when they degrade much more slowly than mRNA~\cite{Shahrezaei2008}.
{In addition, the time-dependent generating function of mRNA is known in closed form~\cite{Iyer-Biswas2009} and can be inverted in some cases to obtain the transient distribution~\cite{Dattani2017}.}

\bigskip

\begin{table}[ht]
\begin{center}
\vspace{-2mm}
\begin{tabular}{ccc}
\hline
Reaction & Rate constant & Interpretation \\
\hline
$G \to G^*$ & $\kon$ & gene activation \\
$G^* \to G$ & $\koff$ & gene inactivation \\
$G^* \to G^* + M$ & $s_0$ & transcription \\
$M \to M + P$ & $s_1$ & translation \\
$M \to \varnothing$ & $d_0$ & mRNA degradation \\
$P \to \varnothing$ & $d_1$ & protein degradation \\
\hline
\end{tabular}
\caption{Chemical reactions defining the two-state model.
The rate constants are usually abbreviated to \emph{rates} as they correspond to actual reactions rates when only one molecule of reactant is present. In the stochastic setting, these rates are in fact propensities, i.e. probabilities per unit of time.}
\label{parameters_single}
\end{center}
\end{table}

In practice, the formulas involve hypergeometric series that are not straightforward to use in a statistical inference framework.
Besides, these series essentially arise from the fact that such a discrete model has to enumerate all potential collisions between molecules (the stochastic mass-action assumption in the master equation).
It is therefore natural to consider keeping only the most important source of noise, that is, keeping a molecular representation for rare species but describing abundant species at a higher level where molecular noise averages out to continuous quantities.
A quick look at reactions in Table~\ref{parameters_single} indicates that the only rare species are $G$ and $G^*$, with quantities $[G]$ and $[G^*]$ being equal to 0 or 1 molecule and satisfying the conservation relation $[G]+[G^*]=1$. The other two, $M$ and $P$, are not conserved quantities in the model and reach a much wider range in biological situations~\cite{Schwanhausser2011}, meaning that saturation constants $s_0/d_0$ and $s_1/d_1$ are much larger than 1 molecule.

\bigskip

Hence, letting $E(t)$, $M(t)$ and $P(t)$ denote the respective quantities of $G^*$, $M$ and $P$ at time $t$, we consider a hybrid version of the previous model, where $E$ has the same stochastic dynamics as before, but with $M$ and $P$ now following usual rate equations:
\begin{equation}%
\label{eq_single_PDMP}
\left\{
\begin{aligned}
{E(t)} &: 0 \xrightarrow[]{\kon} 1, \;\; 1 \xrightarrow[]{\koff} 0 \\
{{M'(t)}} &= s_0 {E(t)} - d_0 {M(t)} \\
{{P'(t)}} &= s_1 {M(t)} - d_1 {P(t)}
\end{aligned}\right.
\end{equation}
This system simply switches between two ordinary differential equations, depending on the value of the two-state continuous-time Markov process $E(t)$, making it a Piecewise-Deterministic Markov Process (PDMP)~\cite{Davis1984}. From a mathematical perspective, model~\eqref{eq_single_PDMP} rigorously approximates the original molecular model when $s_0/d_0$ and $s_1/d_1$ are large enough~\cite{Crudu2009,Crudu2012} and interestingly, it has already been implicitly considered in the biological literature~\cite{Raj2006,Suter2011}.
Note also that the stationary distribution of mRNA is a scaled Beta distribution that is exactly the one of the Beta-Poisson mixture in the discrete model~\cite{Dattani2017}.
Similarly to a recent approach for a two-gene toggle switch~\cite{Lin2016}, we will use~\eqref{eq_single_PDMP} as a basic building block for gene networks.

\bigskip

When both $\kon \ll \koff$ and $d_0 \ll \koff$, mRNA is transcribed by \emph{bursts}, i.e. during short periods which make the mRNA quantity stay far from saturation. Hence, the amount transcribed within each burst is approximately proportional to the burst duration, whose mean is $1/\koff$ by definition: this justifies the quantity $s/\koff$ often being called “burst size” or “burst amplitude”. Furthermore, promoter active periods are much shorter than inactive ones so they can be seen as instantaneous, justifying the name “burst frequency” for the inverse of the mean inactive time $\kon$. We place ourselves in this situation as it often occurs in experiments~\cite{Raj2006,Suter2011,Vinuelas2013,Albayrak2016,Richard2016}.
Note however that these two notions are not clearly defined when relations $\kon \ll \koff$ and $d_0 \ll \koff$ do not hold.

\subsubsection{Adding interactions between genes: the network model}

\begin{figure}[htbp]
\begin{center}
\vspace{-2mm}
\includegraphics[width=\textwidth]{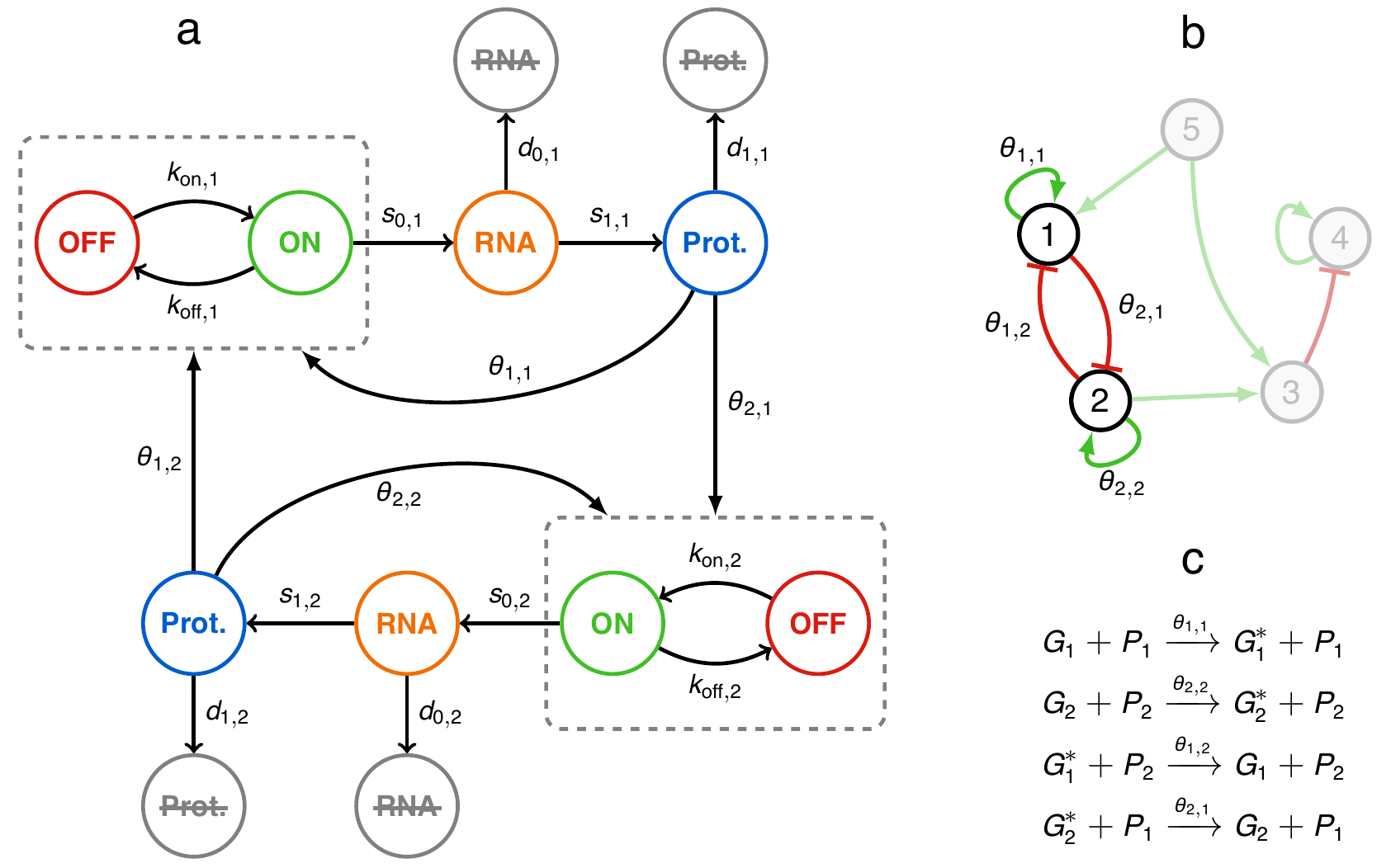}
\caption{(a) Two genes interacting with each other, forming a network. Interactions are assumed to arise from the dependence of promoter dynamics on protein quantities. (b) A higher level of abstraction leads to the traditional gene regulatory network representation. (c) A toy example of reactions defining the interactions between genes $1$ and $2$, making the link between representations (a) and (b).}
\label{scheme_two_state_network}
\end{center}
\end{figure}

Now considering a given set of $n$ genes, a natural way of building a network is to assume that each gene $i$ produces specific mRNA $M_i$ and protein $P_i$, and to define a version of model~\eqref{eq_single_PDMP} with its own parameters:
\begin{equation}%
\label{eq_network_PDMP}
\left\{
\begin{aligned}
E_i(t) &: 0 \xrightarrow[]{\koni} 1, \;\; 1 \xrightarrow[]{\koffi} 0 \\
{M_i}'(t) &= s_{0,i} {E_i(t)} - d_{0,i} {M_i(t)} \\
{P_i}'(t) &= s_{1,i} {M_i(t)} - d_{1,i} {P_i(t)}
\end{aligned}\right.
\end{equation}
Still, genes have static parameters and do not interact with each other. To get an actual network, we need to go one step further: reactions $G_i \to {G_i}^*$ and ${G_i}^* \to G_i$ are not assumed to be elementary anymore, but rather represent complex reactions involving proteins so that promoter parameters $\koni$ and $\koffi$ now depend on proteins~(Fig.~\ref{scheme_two_state_network}a), and {a fortiori} on time.
Our network model will correspond to the explicit definition, for all gene $i$, of functions $\koni(P_1,\dots,P_n)$ and $\koffi(P_1,\dots,P_n)$. These functions shall also depend on network-specific parameters quantifying the interactions, thus making the link between “fitting a chemical model” and “inferring a network”. As a toy example, consider replacing $G_i \to {G_i}^*$ with
two parallel elementary reactions
\begin{equation}
\label{eq_toy_example}
G_i \xrightarrow[]{\theta_{i,0}} {G_i}^* \quad \text{and} \quad G_i + P_j \xrightarrow[]{\theta_{i,j}} {G_i}^* + P_j
\end{equation}
for which applying the law of mass action directly gives $\koni(P_1,\dots,P_n) = \theta_{i,0} + \theta_{i,j} P_j$.
In a regulatory network~(Fig.~\ref{scheme_two_state_network}b), it would correspond to adding a directed edge from gene $j$ to gene $i$, with $\theta_{i,0}$ the basal parameter of gene $i$, and $\theta_{i,j}$ the strength of activation of gene $i$ by protein $P_j$.
We emphasize that the action of $P_j$ on the promoter $G_i$ is not necessarily direct. For example, $P_j$ can instead indirectly modulate the amount/activity of a transcription factor: we suppose in this article that such hidden reactions are fast enough regarding gene expression dynamics so that protein $P_j$ is a relevant proxy for the transcription factor.
Moreover, although we assume here that interactions can only happen at the level of $\koni$ and $\koffi$, mainly for identifiability purposes, it is also possible to make $d_{1,i}$ and $s_{1,i}$ depend on proteins without fundamentally changing the mathematical approach (e.g. see~\cite{Boxma2005,Benaim2012}).

\bigskip

{In order to simplify notations, we normalize model~\eqref{eq_network_PDMP} into a dimensionless equivalent model: we rewrite it in terms of new variables $\overline{M}_i = \frac{d_{0,i}}{s_{0,i}}M_i$ and $\overline{P}_i = \frac{d_{0,i}d_{1,i}}{s_{0,i}s_{1,i}}P_i$, which have values between $0$ and $1$, and report this scale change in the definition of $\koni$ and $\koffi$ (see Appendix~\ref{appendix_timescales} for details). In the remainder of this article, the new variables will still be denoted by $M_i$ and $P_i$ as no confusion arises. The resulting normalized model is:}
\begin{equation}%
\label{eq_network_PDMP_norm}
\left\{
\begin{aligned}
E_i(t) &: 0 \xrightarrow[]{\koni} 1, \;\; 1 \xrightarrow[]{\koffi} 0 \\
{M_i}'(t) &= d_{0,i} \left({E_i(t)} - {M_i(t)}\right) \\
{P_i}'(t) &= d_{1,i} \left({M_i(t)} - {P_i(t)}\right)
\end{aligned}\right.
\end{equation}
still omitting the dependence of $\koni$ and $\koffi$ on $(P_1(t),\dots,P_n(t))$ for clarity. This form enlightens the fact that $s_{0,i}$ and $s_{1,i}$ are just scaling constants: given a path $(E_i,M_i,P_i)_i$ of system~\eqref{eq_network_PDMP_norm}, one can go back to the physical path by simply multiplying $M_i$ by $(s_{0,i}/d_{0,i})$ and $P_i$ by $(s_{0,i}/d_{0,i})\times (s_{1,i}/d_{1,i})$.

\bigskip

Therefore, we get a general network model where each link between two genes is directed and has an explicit biochemical interpretation in terms of molecular interactions. The previous example is very simplistic but one can use virtually any model of chromatin dynamics to derive a form for $\koni$ and $\koffi$, involving hit-and-run reactions, sequential binding, etc.~\cite{Ong2010}. Such aspects are still far from being completely understood~\cite{Coulon2010,Coulon2013,Friedman2015,Bintu2016} and this simple network model can hopefully be used to assess biological hypotheses.
In the next part, we will introduce a more sophisticated interaction form based on an underlying probabilistic model, which is both “statistics-friendly” and interpretable as a non-equilibrium steady state of chromatin environment~\cite{Coulon2013}.

\subsubsection{Some known mathematical results}

Thanks to some recent theoretical results~\cite{Benaim2012,Benaim2015}, simple sufficient conditions on $\koni$ and $\koffi$ ensure that the PDMP network model~\eqref{eq_network_PDMP_norm} is actually well-defined and that the overall joint distribution of $(E_i, M_i, P_i)_i$ converges as $t\to +\infty$ to a unique stationary distribution, which will be the basis of our statistical approach.
Namely, we assume in this article that $\koni$ and $\koffi$ are continuous functions of $(P_1,\dots,P_n)$ and that they are greater than some positive constants.
These conditions are satisfied in most interesting cases, including the above toy example~\eqref{eq_toy_example} when $\theta_{i,0} >0$.

\bigskip

Contrary to creation rates $s_{0,i}$ and $s_{1,i}$, degradation rates $d_{0,i}$ and $d_{1,i}$ play a crucial role in the dynamics of the system. Intuitively, the ratios $(\koni+\koffi)/d_{0,i}$ and $d_{0,i}/d_{1,i}$ respectively control the buffering of promoter noise by mRNA and the buffering of mRNA noise by proteins. A common situation is when promoter and mRNA dynamics are fast compared to proteins, i.e. when $d_{0,i}\gg d_{1,i}$ with $(\koni+\koffi)/d_{0,i}$ fixed. At the limit, the promoter-mRNA noise is fully averaged by proteins and model~\eqref{eq_network_PDMP_norm} simplifies into a deterministic system~\cite{Faggionato2009}:
\begin{equation}%
\label{eq_network_PDMP_ODE}
{P_i}'(t) = d_{1,i} \left(\frac{\koni(\mathbf{P}(t))}{\koni(\mathbf{P}(t)) + \koffi(\mathbf{P}(t))} - {P_i(t)}\right)
\end{equation}
where $\mathbf{P}(t) = (P_1(t),\dots,P_n(t))$. The diffusion limit, which keeps a residual noise, can also be rigorously derived~\cite{Pakdaman2012}. Unsurprisingly, one recovers the traditional way of modelling gene regulatory networks with Hill-type interaction functions.
Equation~\eqref{eq_network_PDMP_ODE} is useful to get an insight into the behaviour of the system~\eqref{eq_network_PDMP_norm} for given $\koni$ and $\koffi$, yet it should be used with caution.
Indeed, the $d_{0,i}/d_{1,i}$ ratio has been shown to span a high range, averaging out to the value $d_{0,i}/d_{1,i}\approx 5$ in mammalian cells~\cite{Schwanhausser2011}, for which taking the limit $d_{0,i}\gg d_{1,i}$ is not obvious. This is consistent with recent single-cell experiments showing a high variability of both mRNA and protein levels between cells~\cite{Albayrak2016}.
In that sense, the PDMP model is much more robust than its deterministic/diffusion counterpart while keeping a similar level of mathematical complexity, which motivates our approach.

\subsubsection{Simulation}

\begin{figure}[ht]
\begin{center}
\includegraphics[width=\textwidth]{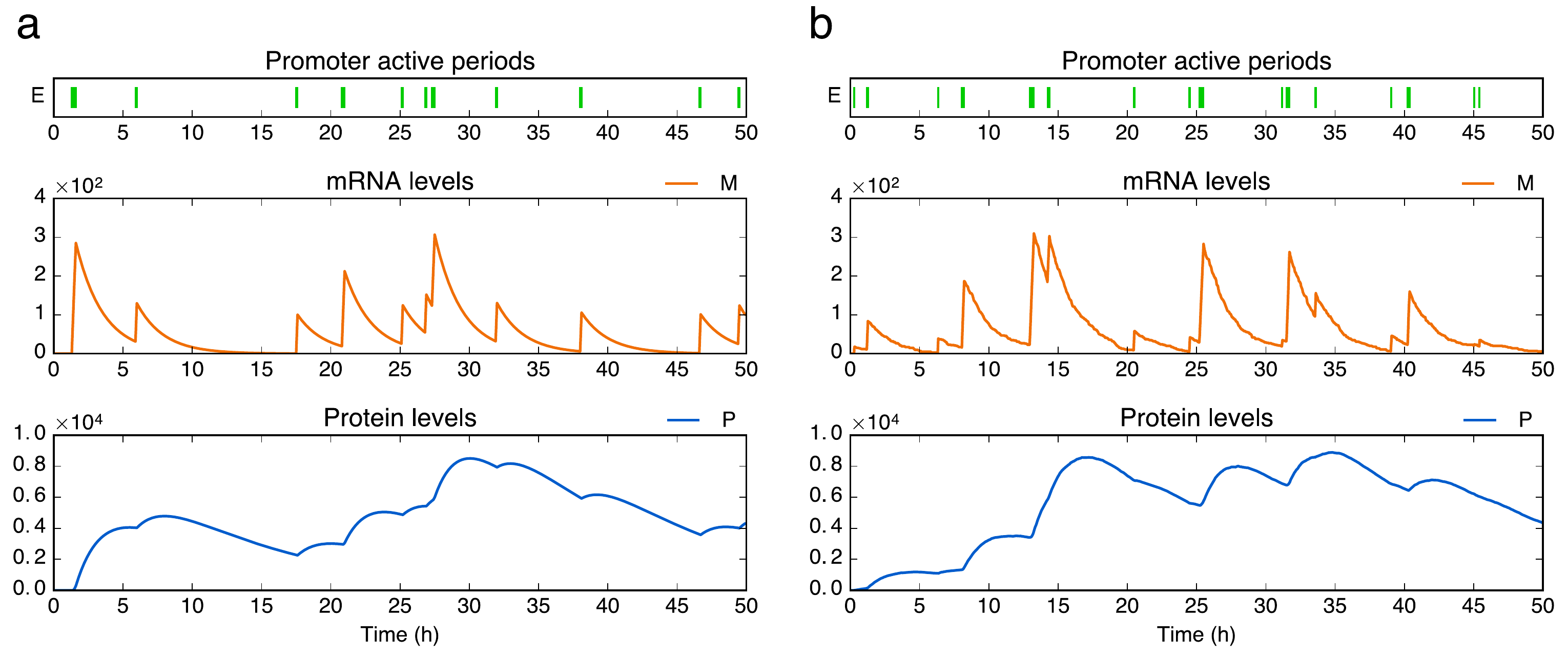}
\caption{Simulations of the two-state model for a single gene. (a) Sample path of the PDMP model using our hybrid numerical scheme (computation time $\approx \SI{0.05}{s}$). (b) Sample path of the classic model using exact stochastic simulation~\cite{Gillespie1977} (computation time $\approx \SI{10}{s}$). Parameters values are $\kon = 0.34$, $\koff = 10$, $s_0 = 10^3$, $s_1 = 10$, $d_0 = 0.5$ and $d_1 = 0.1$ (in $\text{h}^{-1}$).}
\label{sample_path_0}
\end{center}
\end{figure}

We propose a simple algorithm to compute sample paths of our stochastic network model~\eqref{eq_network_PDMP_norm}. It consists in a hybrid version of a basic ODE solver, making it efficient enough to perform massive simulations on large scale networks involving arbitrary numbers of molecules, which would be intractable with a classic molecule-based model (Fig.~\ref{sample_path_0}).
{The deterministic part of the algorithm is a standard explicit Euler scheme, while the stochastic part is based on the transient promoter distribution for single genes: this can be justified by the fact that during a small enough time interval, proteins remain almost constant so genes behave as if $\koni$ and $\koffi$ were constant. We therefore use Bernoulli steps, in a similar way of a diffusion being simulated using gaussian steps.}

\bigskip

After discretizing time with step $\delta t$, the numerical scheme is as follows. Starting from an initial state $({E_i}^{0},{M_i}^{0},{P_i}^{0})_i$, the update of the system from $t$ to $t+\delta t$ is given by:
\begin{equation}%
\label{schema_num_PDMP}
\left\{
\begin{aligned}
{E_i}^{t+\delta t} &\sim \Bb\left(\pi_i^t\right) \\
{M_i}^{t+\delta t} &= (1 - d_{0,i}\delta t){M_i}^{t} + d_{0,i}\delta t {E_i}^{t} \\
{P_i}^{t+\delta t} &= (1 - d_{1,i}\delta t){P_i}^{t} + d_{1,i}\delta t {M_i}^{t}
\end{aligned}\right.
\end{equation}
{where the Bernoulli distribution parameter $\pi_i^t$ is derived by locally solving the master equation for the promoter~\cite{Peccoud1995}, i.e.}
\[\pi_i^t = \frac{a_i^t}{a_i^t + b_i^t} + \left({E_i}^{t} - \frac{a_i^t}{a_i^t + b_i^t}\right)e^{-(a_i^t + b_i^t)\delta t}\]
with the notation $a_i^t = \koni({P_1}^{t},\dots,{P_n}^{t})$ and $b_i^t = \koffi({P_1}^{t},\dots,{P_n}^{t})$.
Intuitively, the algorithm is valid when $\delta t \ll 1/\max_i\left\{{\Koni}, {\Koffi}, {d_{0,i}}, {d_{1,i}} \right\}$ where $\Koni$ and $\Koffi$ denote the maximum values of functions $\koni$ and $\koffi$.

\bigskip

\subsection{Deriving a tractable statistical model}

We will now adopt a statistical perspective in order to deal with gene network inference, considering a set of observed cells. If they are evolving in the same environment for a long enough time, we can reasonably assume that their mRNA and protein levels follow the stationary distribution of an underlying gene network: this distribution can be used as a statistical likelihood for the cells.
Furthermore assuming no cell-cell interactions (which may of course depend on the experimental context), we obtain a standard statistical problem with independent samples.
Since the stationary distribution of the stochastic network model~\eqref{eq_network_PDMP_norm} is well-defined but {a priori} not analytically tractable, we will derive an explicit approximation and then reduce our inference problem to a traditional likelihood-based estimation. We will do so in two cases: when there is no self-interaction, and for a specific form of auto-activation.

\subsubsection{Separating mRNA and protein timescales}

It is for the moment very rare to experimentally obtain the amount of proteins for many genes at the single-cell level. We will therefore assume here that only mRNAs are observed. To deal with this problem, we take the protein timescale as our reference by fixing $d_{1,i}$ and assume that promoter dynamics are faster than proteins, i.e. $(\koni+\koffi) \gg d_{1,i}$ in a biologically relevant way, say $(\koni+\koffi)/d_{1,i} > 10$ (thus the deterministic limit~\eqref{eq_network_PDMP_ODE} does not necessarily hold).
{Furthermore, in line with several recent experiments~\cite{Li2011,Albayrak2016}, we assume that $d_{0,i}$ is sufficiently larger than $d_{1,i}$ so that the correlation between mRNAs and proteins produced by the gene is very small: model~\eqref{eq_network_PDMP_norm} then can be reduced by removing mRNA and making proteins directly depend on the promoters (see Appendix~\ref{appendix_timescales}).} The result is
\begin{equation}%
\label{eq_network_PDMP_red}
\left\{
\begin{aligned}
E_i(t) &: 0 \xrightarrow[]{\koni} 1, \;\; 1 \xrightarrow[]{\koffi} 0 \\
{P_i}'(t) &= d_{1,i} \left({E_i(t)} - {P_i(t)}\right)
\end{aligned}\right.
\end{equation}
which still admits the deterministic limit~\eqref{eq_network_PDMP_ODE}. Since mRNA dynamics are faster than proteins, one can also assume that, given protein levels $\mathbf{P} = (P_1,\dots,P_n)$, each mRNA level $M_i$ follows the quasi-steady state distribution
\begin{equation}%
\label{eq_RNA_beta}
M_i \,|\, \mathbf{P} \sim \Bet\left(\frac{\koni(\mathbf{P})}{d_{0,i}},\frac{\koffi(\mathbf{P})}{d_{0,i}}\right)
\end{equation}
corresponding to the single-gene model~\cite{Boxma2005,Dattani2017} with constant parameters $\koni(\mathbf{P})$ and $\koffi(\mathbf{P})$. Numerically, this approximation works well even for moderate values of $d_{0,i}$, such as $d_{0,i} = 5 \times d_{1,i}$ (see the Results section).

\bigskip

Biologically, equations~\eqref{eq_network_PDMP_red} and~\eqref{eq_RNA_beta} suggest that correlations between mRNA levels may not directly arise from correlations between promoters \emph{states} (which in fact are weak because of $(\koni+\koffi) \gg d_{1,i}$), but rather originate from correlations between promoter \emph{parameters} $\koni$ and $\koffi$, which themselves depend on the protein joint distribution.

\bigskip

Table~\ref{table_models} sums up the successive modelling steps introduced so far.
From now on, we will always assume the form~\eqref{eq_RNA_beta} for the mRNA distribution, and thus our model is reduced to equation~\eqref{eq_network_PDMP_red} which only involves proteins.

\bigskip

\begin{table}[ht]
\begin{center}
\begin{tabular}{|c|l|cl}
\cline{1-2}
\multirow{3}{*}{1} & \textbf{Single-gene, discrete}~\cite{Shahrezaei2008} & & \\
& {\small $\diamond$ All molecules are discrete} & & \\
& {\small $\diamond$ mRNA distribution: Beta-Poisson} & \multirow{2}{*}{$\downarrow$} & \hspace{-2mm}\multirow{2}{4cm}{Abundant species treated continuously} \\
\cline{1-2}
\multirow{3}{*}{2} & \textbf{Single-gene, PDMP}~\eqref{eq_single_PDMP} & & \\
& {\small $\diamond$ Only the promoter is discrete} & & \\
& {\small $\diamond$ mRNA distribution: Beta} & \multirow{2}{*}{$\downarrow$} & \hspace{-2mm}\multirow{2}{4cm}{Introduction of interactions via $\kon$, $\koff$} \\
\cline{1-2}
\multirow{3}{*}{3} & \textbf{Network}~\eqref{eq_network_PDMP}\textbf{, normalized version}~\eqref{eq_network_PDMP_norm} & & \\
& {\small $\diamond$ Both accurate and fast to simulate} & & \\
& {\small $\diamond$ mRNA distribution: unknown} & \multirow{2}{*}{$\downarrow$} & \hspace{-2mm}\multirow{2}{4.5cm}{Timescale separation of Protein/mRNA ($d_0 \gg d_1$)} \\
\cline{1-2}
\multirow{3}{*}{4} & \textbf{Simplified network}~\eqref{eq_network_PDMP_red} & & \\
& {\small $\diamond$ mRNA is removed from the network} & & \\
& {\small $\diamond$ Conditional mRNA distribution: Beta~\eqref{eq_RNA_beta}}& & \\
\cline{1-2}
\end{tabular}
\caption{Successive dynamical models introduced in this article, recalling for each step the main feature and the form of the mRNA stationary distribution. The full network model (step 3) is used for simulations, while the simplified one (step 4) is used to derive the approximate statistical likelihood.}
\label{table_models}
\end{center}
\end{table}

\subsubsection{Hartree approximation}

In this section, we present the Hartree approximation principle and provide an explicit formula in the particular case of no self-interaction.
The simplified model~\eqref{eq_network_PDMP_red} is still not analytically tractable, but it is now appropriate for employing the \emph{self-consistent proteomic field} approximation introduced in~\cite{Sasai2003,Walczak2005} and successfully applied in~\cite{Kim2007,Zhang2014}.
More precisely, we will use its natural PDMP counterpart, which will be referred to as “Hartree approximation” since the main idea is similar to the Hartree approximation in physics~\cite{Sasai2003}.
It consists in assuming that genes behave as if they were independent from each other, but submitted to a common “proteomic field” created by all other genes. In other words, we transform the original problem of dimension $2^n$ into $n$ independent problems of dimension $2$ that are much easier to solve (see Appendix~\ref{appendix_hartree} for details).

\bigskip

When $\koni$ and $\koffi$ do not depend on $P_i$ (i.e. no self-interaction), this approach results in approximating the protein stationary distribution of model~\eqref{eq_network_PDMP_red} by the function
\begin{equation}%
\label{eq_hartree_prot}
u(y) = \prod_{i=1}^n \frac{{y_i}^{a_i(y)-1} {(1-y_i)}^{b_i(y)-1}}{\Beta(a_i(y),b_i(y))}
\end{equation}
where $y = (y_1,\dots,y_n) = (P_1,\dots,P_n) = \mathbf{P}$, $a_i(y) = {\koni(y)}/{d_{1,i}}$, $b_i(y) = {\koffi(y)}/{d_{1,i}}$ and $\Beta$ is the standard Beta function. Note that promoter states have been integrated out since they are not required by equation~\eqref{eq_RNA_beta}.

\bigskip

The function $u$ is a heuristic approximation of a probability density function. It is only valid when interactions are not too strong, that is, when $\koni$ and $\koffi$ are close enough to constants, and it becomes exact when they are true constants.
Besides, it does not integrate to $1$ in general.
However, this approximation turns out to be very robust in practice and it has the great advantage to be fully explicit (and significantly simpler than in the non-PDMP case), thus providing a promising base for a statistical model.

\bigskip

When $\koni$ and $\koffi$ depend on $P_i$, one can still explicitly compute the Hartree approximation in many cases: we will give an example in the next section.
Alternatively, it is always possible to use formula~\eqref{eq_hartree_prot} even with self-interactions, giving a correct approximation when the feedback is not too strong, as for other proteins.

\subsubsection{An explicit form for interactions}

We now propose an explicit definition of functions $\koni$ and $\koffi$. Recent work~\cite{Vinuelas2013,Senecal2014,Fukaya2016} showed that apparent increased transcription actually reflects an increase in burst frequency rather than amplitude. We therefore decided to model only $\koni$ as an actual function and to keep $\koffi$ constant. In this view, the activation frequency of a gene can be influenced by ambiant proteins, whereas the active periods have a random duration that is dictated only by an intrinsic stability constant of the transcription machinery.

\bigskip

Our approach uses a description of the molecular activity around the promoter in a very similar way as Coulon et al.~\cite{Coulon2010}. Accordingly, we make a quasi-steady state assumption to obtain $\koni$. This idea based on thermodynamics was also used in the DREAM3 in-Silico Challenge~\cite{Marbach2010} to simulate gene networks. However, only mean transcription rate was described (instead of promoter activity in our work), which is inappropriate to model bursty mRNA dynamics at the single-cell level.

\bigskip

We herein derive $\koni$ from an underlying stochastic model for chromatin dynamics. We first introduce a set of abstract chromatin states, each state being associated with one of two possible rates of promoter activation, either a low rate $k_{0,i}$ or a high rate $k_{1,i} \gg k_{0,i}$.
More specifically, such chromatin states may be envisioned as a coarse-grained description of the chromatin-associated parameters that are critical for transcription of gene $i$.
Second, we assume a separation of timescales between the abstract chromatin model and the promoter activity, so that the promoter activation reaction depends only on the quasi-steady state of chromatin. In other words, the effective $\koni$ is a combination of $k_{0,i}$ and $k_{1,i}$ which integrates all the chromatin states: its value depends on the probability of each state and {a fortiori} on the transitions between them.
We propose a transition scheme which leads to an explicit form for $\koni$, based on the idea that proteins can alter chromatin by hit-and-run reactions and potentially introduce a memory component. Some proteins thereby tend to stabilize it either in a “permissive” configuration (with rate $k_{1,i}$) or in a “non-permissive” configuration (with rate $k_{0,i}$), providing notions of \emph{activation} and \emph{inhibition}.
A more precise definition and details of the derivation are provided in Appendix~\ref{appendix_interactions}.

\bigskip

\begin{figure}[ht]
\begin{center}
\includegraphics[width=\textwidth]{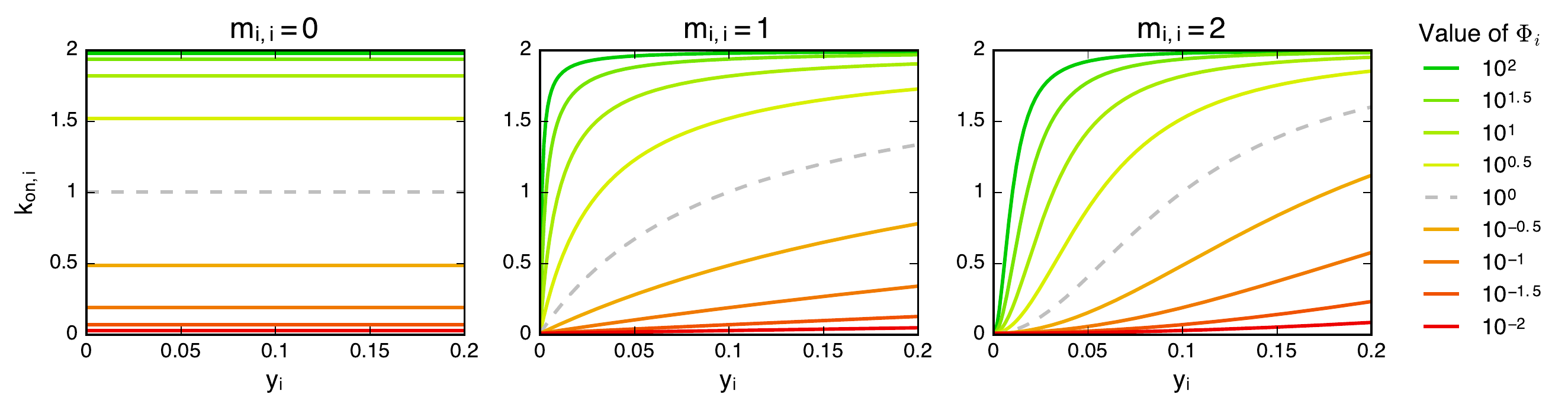}
\caption{Different auto-activation types in the network model. Each color corresponds to a fixed value of $\Phi_i$ in formula~\eqref{eq_kon_full}, and each curve represents $\koni$ as a function of $y_i$ for $m_{i,i} = 0$ (no feedback), $m_{i,i} = 1$ (monomer-type feedback) and $m_{i,i} = 2$ (dimer-type feedback).
The neutral value $\Phi_i = 1$ is represented by a dashed gray line.
Here $k_{0,i} = 0.01$, $k_{1,i} = 2$ and $s_{i,i} = 0.1$.}
\label{interaction_curves}
\end{center}
\end{figure}

The final form is the following. First, we define a function of every protein but $P_i$,
\[\Phi_i(y) = \exp(\theta_{i,i}) \prod_{j\neq i} \frac{1 + \exp(\theta_{i,j}) (y_j/s_{i,j})^{m_{i,j}}}{1 + (y_j/s_{i,j})^{m_{i,j}}}\]
which may represent the external input of gene $i$.
Then, $\koni$ is defined by
\begin{equation}%
\label{eq_kon_full}
\koni(y) = \frac{k_{0,i} + k_{1,i} \Phi_i(y)(y_i/s_{i,i})^{m_{i,i}}}{1 + \Phi_i(y)(y_i/s_{i,i})^{m_{i,i}}} .
\end{equation}
Hence, when the input $\Phi_i(y)$ is fixed, $\koni$ is a standard Hill function which describes how gene~$i$ is self-activating, depending on the Hill coefficient $m_{i,i}$ (Fig.~\ref{interaction_curves}). The neutral value is set to $\Phi_i(y) = 1$, so that for this particular value, $s_{i,i}$ is the usual dissociation constant. Moreover, if $\theta_{i,j} = 0$ for all $j\neq i$, then $\Phi_i$ becomes the constant function $\Phi_i(y) = \exp(\theta_{i,i})$, and thus $\theta_{i,i}$ may be seen as a “basal” parameter, summing up all potential hidden inputs. On the contrary, if some $\theta_{i,j} > 0$ (resp. $\theta_{i,j} < 0)$, then $\Phi_i$ becomes itself an increasing (resp. decreasing) Hill-type function of protein $P_j$, where $m_{i,j}$ and $s_{i,j}$ again play their usual roles.

\bigskip

The $n \times n$ matrix $\theta = (\theta_{i,j})$ therefore plays the same role as the interaction matrix in traditional network inference frameworks~\cite{Mizeranschi2015}. For $i\neq j$, $\theta_{i,j}$ quantifies the regulation of gene $i$ by gene $j$ (activation if $\theta_{i,j}>0$, inhibition if $\theta_{i,j}<0$, no influence if $\theta_{i,j} = 0$), and the diagonal term $\theta_{i,i}$ aggregates the “basal input” and the “self-activation strength” of gene $i$. Note that self-inhibition could be considered instead, but the choice has to be made before the inference since the self-interaction form is notoriously difficult to identify, especially in the stationary regime. In the remainder of this article, we assume that parameters $k_{0,i}$, $k_{1,i}$, $m_{i,j}$ and $s_{i,j}$ are known and we focus on inferring the matrix $\theta$.

\bigskip

A benefit of the interaction form~\eqref{eq_kon_full} is to allow for a fully explicit Hartree approximation of the protein distribution (see Appendix~\ref{appendix_interactions} for details). In particular, if $m_{i,i} > 0$ and $c_i = (k_{1,i} - k_{0,i})/(d_{1,i} m_{i,i})$ is a positive integer for all $i$, the approximation is given by
\begin{equation}%
\label{eq_hartree_prot_model}
u(y) = \prod_{i=1}^n \left(\sum_{r=0}^{c_i} p_{i,r}(y) \frac{{y_i}^{a_{i,r}-1} {(1-y_i)}^{b_i-1}}{\Beta(a_{i,r},b_i)} \right)
\end{equation}
with $a_{i,r} = ((c_i - r)k_{0,i} + rk_{1,i})/(d_{1,i}c_i)$, $b_i = \koffi/d_{1,i}$ and
\[p_{i,r}(y) = \frac{\binom{c_i}{r} \Beta(a_{i,r},b_i) (\Phi_i(y)/s_{i,i}^{m_{i,i}})^r}{\sum_{r' = 0}^{c_i} \binom{c_i}{r'} \Beta(a_{i,r'},b_i) (\Phi_i(y)/s_{i,i}^{m_{i,i}})^{r'}} .\]
In other words, the Hartree approximation~\eqref{eq_hartree_prot_model} is a product of gene-specific distributions which are themselves mixtures of Beta distributions: for gene $i$, the $a_{i,r}$ correspond to “frequency modes” ranging from $k_{0,i}$ to $k_{1,i}$, weighted by the probabilities $p_{i,r}(y)$. It is straightforward to check that inhibitors tend to select the low burst frequencies of their target ($a_{i,r} \approx k_{0,i}$) while activators select the high frequencies ($a_{i,r} \approx k_{1,i}$).
If $m_{i,i} = 0$ for some $i$, then $\koni$ does not depend on $P_i$ so one just has to replace the $i$-th term in the product~\eqref{eq_hartree_prot_model} with the single Beta form as in equation~\eqref{eq_hartree_prot}, which is equivalent to taking the limit $c_i \to +\infty$.
Finally, when $m_{i,i} > 0$ but $c_i$ is not an integer, using $\lceil c_i \rceil$ instead keeps a satisfying accuracy.

\subsubsection{The statistical model in practice}

Our statistical framework simply consists in combining the timescale separation~\eqref{eq_RNA_beta} and the Hartree approximation~\eqref{eq_hartree_prot_model} into a standard {hidden Markov model}. Indeed, conditionally to the proteins, mRNAs are independent and follow well-defined Beta distributions
\begin{equation}%
\label{eq_hartree_RNA}
v(x,y) = \prod_{i=1}^n \frac{{x_i}^{\widetilde{a}_i(y)-1} {(1-x_i)}^{\widetilde{b}_i(y)-1}}{\Beta(\widetilde{a}_i(y),\widetilde{b}_i(y))}
\end{equation}
where $x = (x_1,\dots,x_n) = (M_1,\dots,M_n) = \mathbf{M}$, $\widetilde{a}_i(y) = {\koni(y)}/{d_{0,i}}$ and $\widetilde{b}_i(y) = {\koffi(y)}/{d_{0,i}}$. Then one can use~\eqref{eq_hartree_prot_model} to approximate the joint distribution of proteins.
Hence, recalling the unknown interaction matrix $\theta$, the inference problem for $m$ cells with respective levels $(\mathbf{M}_k,\mathbf{P}_k)_{1\leqslant k \leqslant m}$ is based on the (approximate) complete log-likelihood:
\begin{equation}%
\label{eq_likelihood}
\begin{aligned}
\ell &= \ell(\mathbf{M}_1,\dots,\mathbf{M}_m, \mathbf{P}_1,\dots,\mathbf{P}_m | \theta) = \sum_{k=1}^m \log(u(\mathbf{P}_k)) + \log(v(\mathbf{M}_k,\mathbf{P}_k))
\end{aligned}
\end{equation}
where we used conditional factorization and independence of the cells.

\bigskip

The basic statistical inference problem would be to maximize the marginal likelihood of mRNA with respect to $\theta$. Since this likelihood has no simple form, a typical way to perform inference is to use an Expectation-Maximization (EM) algorithm on the complete likelihood~\eqref{eq_likelihood}.
However, the algorithm may be slow in practice because of the computation of expectations over proteins. A faster procedure consists in simplifying these expectations using the distribution modes: the resulting algorithm is often called “hard EM” or “classification EM” and is used in the Results section.
Moreover, it is possible to encode some potential knowledge or constraints on the network by introducing a prior distribution $w(\theta)$. In this case, from Baye's rule, one can perform \emph{maximum a posteriori} (MAP) estimation of $\theta$ by using the same EM algorithm but adding the penalization term $\log(w(\theta))$ to $\ell$ during the Maximization step (see Appendix~\ref{sec_EM_MAP} and the Results section).
Alternatively, a full bayesian approach, i.e. sampling from the posterior distribution of $\theta$ conditionally to $(\mathbf{M}_1,\dots,\mathbf{M}_m)$, may also be considered using standard MCMC methods.

\bigskip

Taking advantage of the latent structure of proteins, we can also deal with missing data in a natural way: if the mRNA measurement of gene $i$ is invalid in a cell $k$ owing to technical problems, it is possible to ignore it by removing the $i$-th term in the conditional distribution of mRNAs~\eqref{eq_hartree_RNA}. This only modifies the definition of $v$ for cell $k$ in equation~\eqref{eq_likelihood}, ensuring that all valid data is effectively used for each cell.

\section{Results}

In this part, we first compare the distribution of the mechanistic model~\eqref{eq_network_PDMP_norm} to the mRNA quasi-steady state combined with Hartree approximation for proteins, on a simple toggle-switch example. Then, we show that the single-gene model with auto-activation can fit marginal mRNA distributions from real data better than the constant-$\kon$ model. Finally, we successfully apply the inference procedure to various two-gene networks simulated from the mechanistic model.

\subsection{Relevance of the approximate likelihood}

\begin{figure}[ht]
\begin{center}
\includegraphics[width=0.9\textwidth]{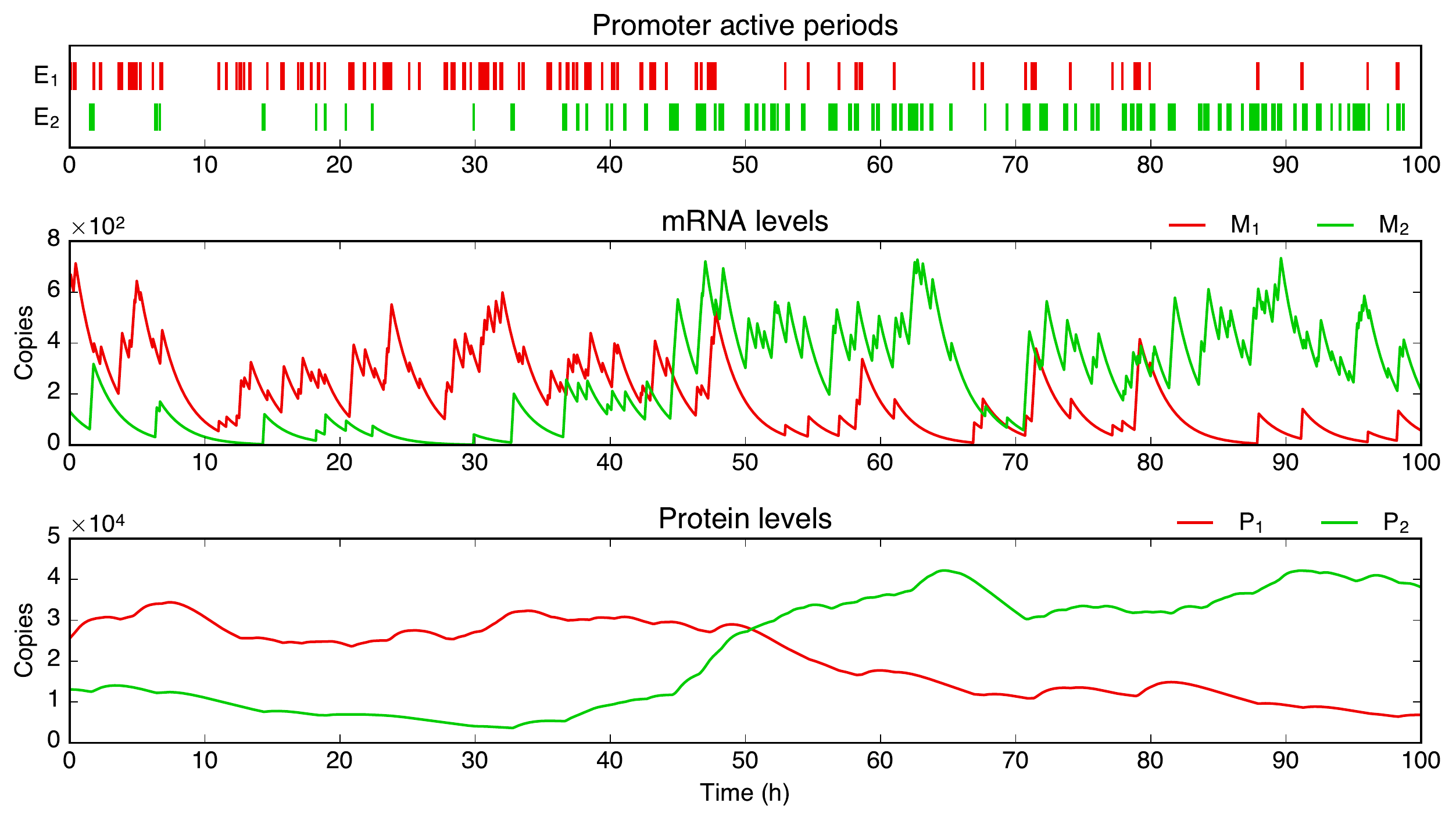}
\caption{Sample path of a two-gene toggle switch, with gene 1 in red and gene 2 in green. While always staying in a bursty regime regarding mRNAs, genes can switch between high and low frequency modes (here at $t\approx 50\si{h}$). From this example, it is clear that the overall joint distribution can contain correlations even if the bursts themselves are not coordinated.}
\label{sample_path_1}
\end{center}
\end{figure}

\begin{figure}[ht]
\begin{center}
\includegraphics[width=0.71\textwidth]{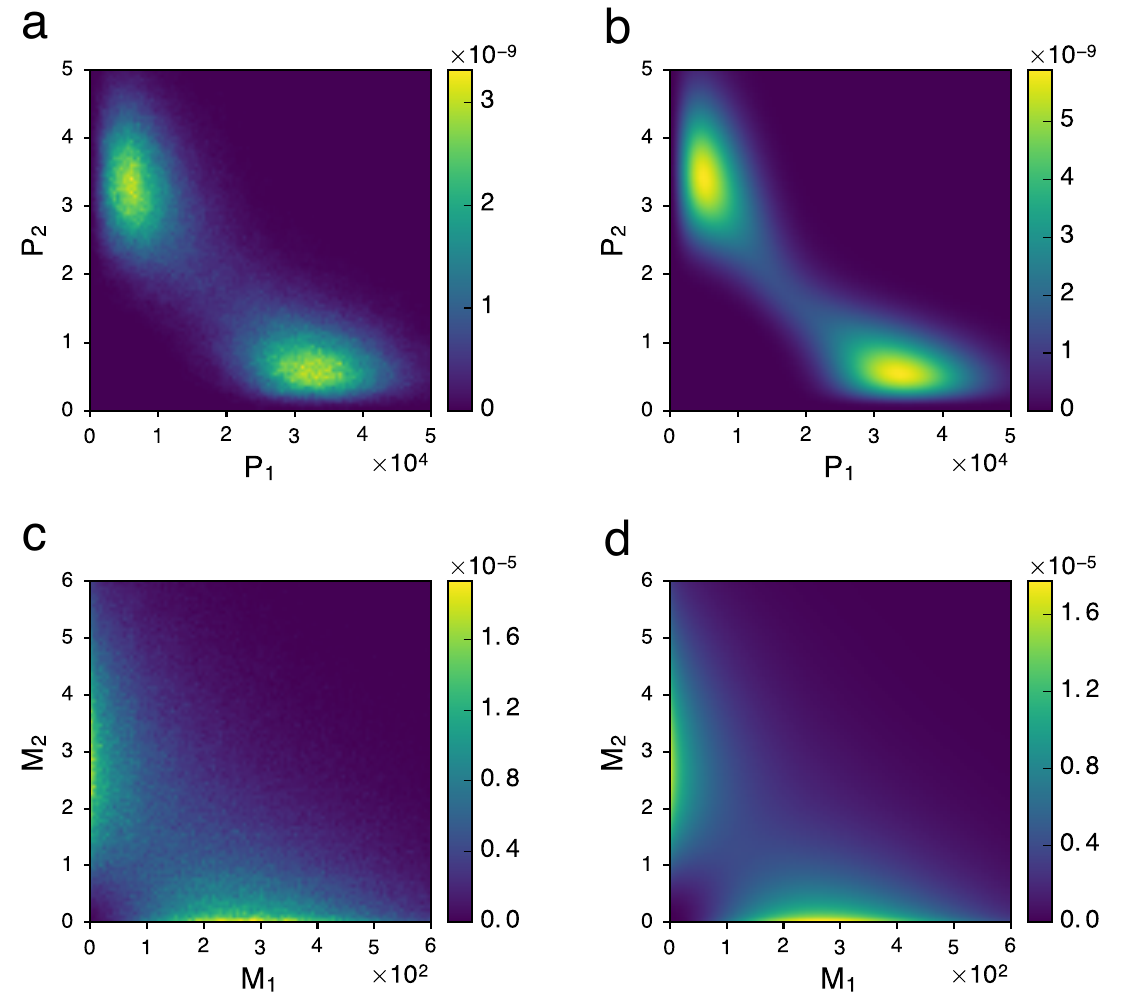}
\caption{Exact and approximate stationary distributions for the example of toggle switch. True distributions (left side) were estimated by sample path simulation, while approximations (right side) have explicit formulas. (a) True distribution of proteins. (b) Approximate distribution of proteins, from formula~\eqref{eq_hartree_prot_model}. (c) True distribution of mRNAs. (d) Approximate distribution of mRNAs, obtained by integrating the conditional distribution of mRNA~\eqref{eq_hartree_RNA} against (b).}
\label{Exemple_Hartree_1}
\end{center}
\end{figure}

Starting from the normalized mechanistic model~\eqref{eq_network_PDMP_norm}, two approximations were used to derive the final statistical likelihood~\eqref{eq_likelihood}: the quasi-steady state assumption for mRNAs given protein levels, and the Hartree approximation for the joint distribution of proteins. Crucially, this approximate likelihood has to be close enough to the exact one in order to preserve the equivalence between inferring a network and fitting the mechanistic model.
To get an idea of the accuracy, we considered a basic two-gene toggle switch defined by $\koni$ following equation~\eqref{eq_kon_full} with the interaction matrix given by $\theta_{1,1} = \theta_{2,2} = 4$ and $\theta_{1,2} = \theta_{2,1} = -8$ (full parameter list in Appendix~\ref{appendix_parameters}). By computing sample paths (Fig.~\ref{sample_path_1}), we estimated the stationary distribution and compared it with our approximation, which appeared to be very satisfying, both for proteins and mRNAs (Fig.~\ref{Exemple_Hartree_1}).

\subsection{Fitting marginal mRNA distributions from real data}

A particularity of single-cell data is to often exhibit bursty regimes for mRNA (meaning $\kon \ll \koff$ and $d_0 \ll \koff$) and potentially also for proteins (adding $d_1 \ll \koff$), which are well fitted by Gamma distributions~\cite{Albayrak2016}.
At this stage, it is worth mentioning that the Gamma distribution can be seen as a limit case of the Beta distribution. Intuitively, when $b \gg 1$ and $b \gg a$ (typically $a = \kon/d_0$ and $b = \koff/d_0$), most of the mass of the distribution $\Bet(a,b)$ is located at $x\ll 1$ so we have the first order approximation
\[ \begin{aligned}
x^{a-1}(1-x)^{b-1} &= x^{a-1}\exp((b-1)\log(1-x)) \\
& \approx x^{a-1}\exp(-bx)
\end{aligned}\]
and thus $\Bet(a,b) \approx \gamma(a,b)$. This way, formulas~\eqref{eq_hartree_prot_model} and~\eqref{eq_hartree_RNA} can be easily transformed into Gamma-based distributions.
Parameters $s_0$ and $\koff$ then aggregate in $\koff/s_0$ because of the scaling property of the Gamma distribution, so only this ratio has to be inferred: from an applied perspective, it simply represents a scale parameter for each gene. This remark leads to a possible preprocessing phase that can be used for estimating the crucial basal parameters of the network, without requiring the knowledge of such scale parameters (see Appendix~\ref{appendix_data}).

\bigskip

\begin{figure}[ht]
\begin{center}
\includegraphics[width=0.45\textwidth]{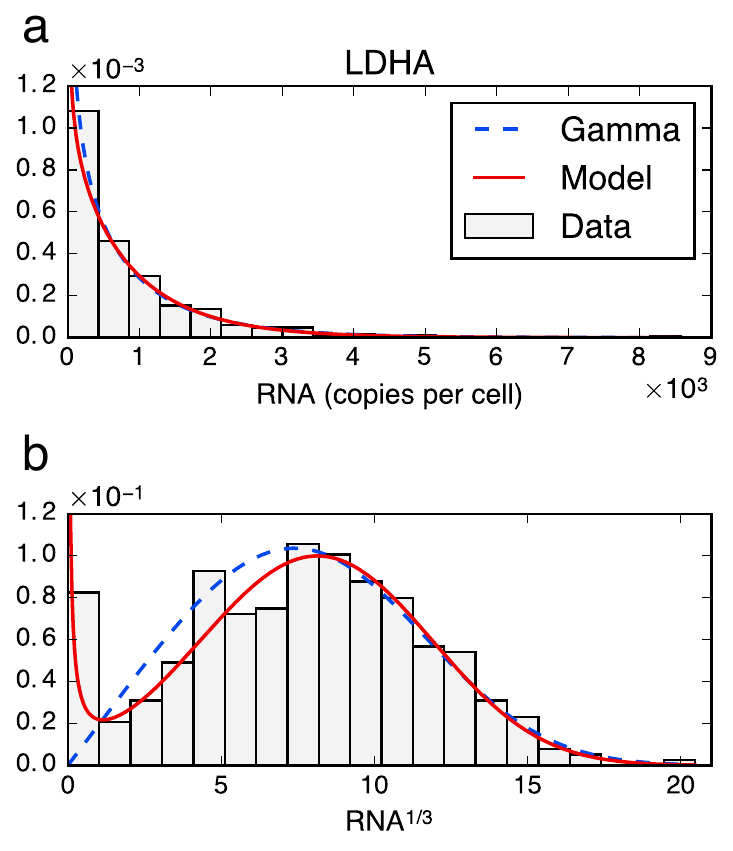}
\caption{Fitting marginal distributions from real single-cell data: example of the LDHA gene. The red curve is the stationary distribution associated with our interaction form (here a single gene with auto-activation), while the dashed blue curve corresponds to the basic two-state model in the bursty regime (Gamma distribution). (a) The raw data seems to be well fitted by the Gamma distribution, which in this view is close to our model. (b) Same fit viewed after applying the transformation $x\mapsto x^{1/3}$. The data becomes bimodal and the fit appears to be better with the auto-activation model.}
\label{data_fitting}
\end{center}
\end{figure}

{In addition, our network model is able to generate multiple modes while keeping such bursty regimes (Fig.~\ref{sample_path_1}), as noticeable in the stationary distribution~\eqref{eq_hartree_prot_model}. Interestingly, this feature has already been considered in the literature by empirically introducing mixture distributions~\cite{Gu2015,Ghazanfar2016}. As a first step toward applications, we compared our model in the simplest case (independent genes with auto-activation) to marginal distributions of single-cell mRNA measurements from~\cite{Richard2016}. Our model was fitted and compared to the basic two-state model in the bursty regime, i.e. to a simple Gamma distribution: Fig.~\ref{data_fitting} shows the example of the LDHA gene.
Although very close when viewed in raw molecule numbers, the distributions differ after applying the transformation $x\mapsto x^\alpha$ with $\alpha = 1/3$, which tends to compress great values while preserving small values. The data becomes bimodal, suggesting the presence of two bursting regimes, a “normal” one and a very small “inhibited” one: the auto-activation model then performs better than the simple Gamma, which necessarily stays unimodal for $0<\alpha<1$.
Note that the RTqPCR protocol used in~\cite{Richard2016} was shown to be far more sensitive than single-cell RNA-seq in the detection of low abundance transcripts~\cite{Mojtahedi2016}. Since the data also contains small nonzero values, this tends to support a true biological origin for the peak in zero. Besides, the case of distributions that are not bimodal until transformed also arises for proteins~\cite{Sokolik2015}.}

\subsection{Application of the inference procedure}

By construction of the mechanistic model, the interaction matrix $\theta$ can describe any oriented graph by explicitly defining causal quantitative links between genes, which is difficult to do within traditional statistical frameworks (e.g. bayesian networks or undirected Markov random fields). The logical downside is that identifiability issues seem inevitable.
In a first attempt to assess this aspect, we implemented the inference method presented above and tested it on various two-gene networks, assuming auto-activation for each gene (i.e. $m_{i,i} > 0$) with equation~\eqref{eq_kon_full} to maximize variability without considering perturbations of the system (parameter list in Appendix~\ref{appendix_parameters}).

\bigskip

We decided to investigate the worst case scenario in terms of cell numbers. We are fully aware of the existence of technologies allowing to interrogate thousands of cells simultaneously, but most of the recent studies still rely upon a much smaller number of cells.
For each network, we therefore simulated mRNA snapshot data for 100 cells using the full PDMP model~\eqref{eq_network_PDMP_norm}. We then inferred the matrix $\theta$ using a “hard EM” algorithm based on the likelihood~\eqref{eq_likelihood}, that is, alternatively maximizing the likelihood with respect to $\theta$ and with respect to the (unknown) protein levels of each cell. A lasso-like penalization term, corresponding to a prior distribution, was added to the $\theta_{i,j}$ for $i\neq j$ to obtain true zeros -- so that the inferred network topology is clear -- and to prevent keeping both $\theta_{i,j}$ and $\theta_{j,i}$ when one is significantly weaker (see Appendix~\ref{sec_EM_MAP} for details of the penalization and the whole procedure).

\bigskip

\begin{figure}[!ht]
\begin{center}
\includegraphics[width=0.82\textwidth]{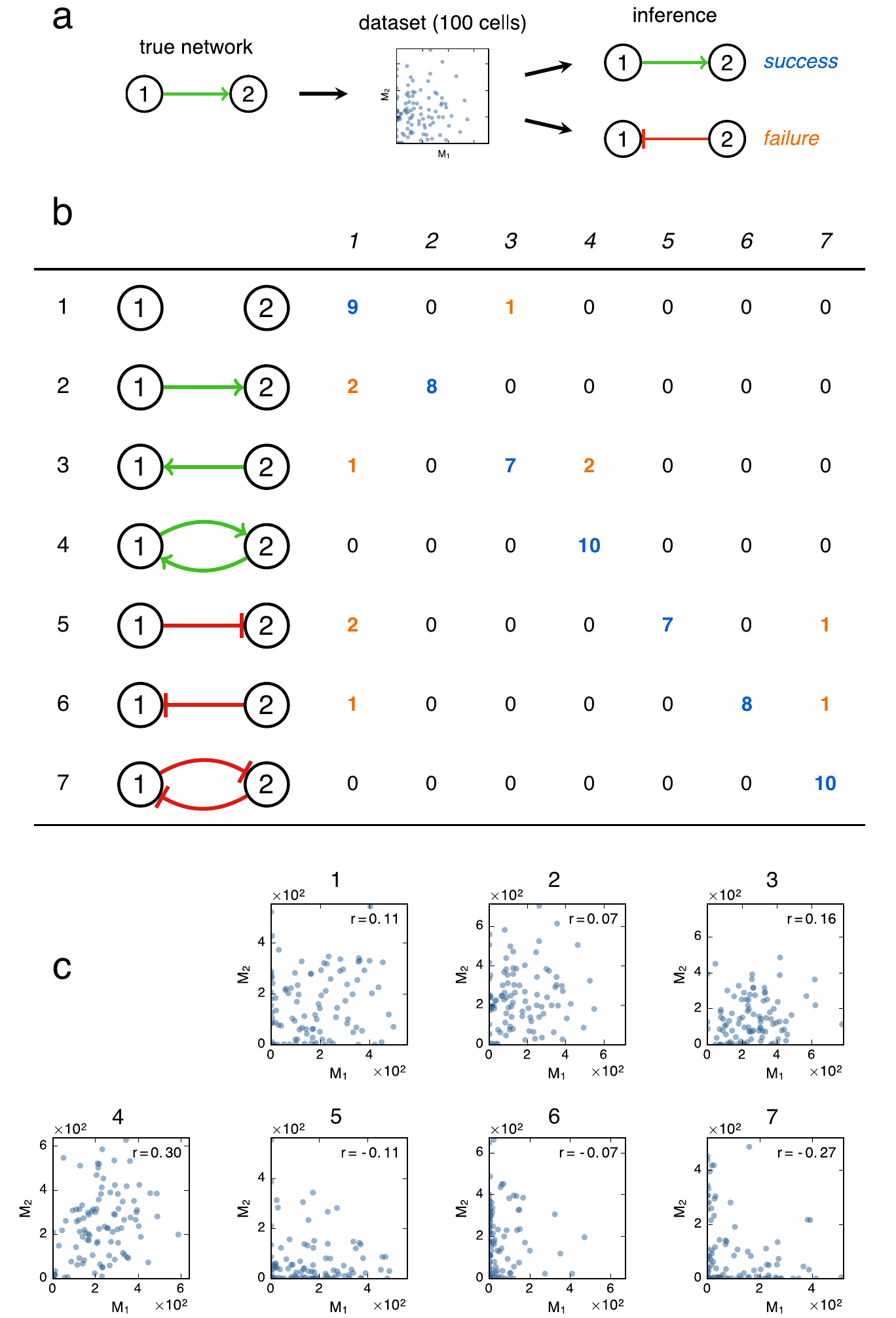}
\caption{Testing our inference method on simple networks. (a) For each network, numbered from 1 to 7, we simulated 100 cells using the full mechanistic model until the stationary regime was reached. Then we took a snapshot of their mRNA levels and inferred the parameters from this data. The result was called successful when the inferred structure (topology and nature of the links) was the same as the true network.
(b) For each network (rows), 10 datasets were simulated and the results were reported by counting the number of inferred $\theta$ corresponding to each structure (columns), highlighting successes (blue) and failures (orange). The perfect inference would lead to 10 for all the diagonal terms and 0 everywhere else. (c) Examples of simulated mRNA datasets (one for each network). Although having coherent signs, Pearson's correlation coefficients (top right of each plot) would clearly be insufficient to distinguish between the different networks.}
\label{test_networks}
\end{center}
\vspace{-8mm}
\end{figure}

We obtained highly encouraging results since every structure was inferred with a high probability of success (Fig.~\ref{test_networks}), meaning that the non-diagonal (i.e. interaction) terms of $\theta$ had the right sign and were nonzero at the right places. A list of the inferred values is provided in Table~\ref{tab_inferred_networks} of Appendix~\ref{appendix_parameters}. {It is very important at that stage to emphasize that we are not trying to infer $\theta$ exactly: we only assess whether it has a zero or nonzero value and its sign.}
Although the results tend to support the identifiability of the full matrix $\theta$ in this simple two-gene case, one has to be aware that the quantity we maximize (an approximate likelihood) is a priori nonconvex and can have several local maxima (i.e. networks that are relevant candidates to explain the data). The result of the inference thus can depend on the starting point: in this first approach we chose the null matrix to be the starting point for $\theta$, which corresponds to the -- biologically relevant -- expectation of “balanced” behaviors (e.g. we do not expect $\theta_{1,1} \ll \theta_{2,2}$). Alternatively, one can consider some probabilistic prior knowledge on $\theta$ to implement a (possibly rough) idea of parameter values from a Bayesian viewpoint: it is worth mentioning that any knockout information can be implemented this way in our model.

\bigskip

Finally, we assessed the inference behavior in the presence of dropouts, i.e. genes expressed at a low level in a cell that give rise to zeros after measurement~\cite{Wagner2016}. Our first tests tend to indicate that our approach is robust regarding dropouts, in the sense that up to 30\% of simulated dropouts does not drastically affect the estimation of $\theta$ once the other parameters have been estimated correctly (see Table~\ref{tab_inferred_networks_dropout} of Appendix~\ref{appendix_parameters} for an example).


\section{Discussion}

In this paper, we introduce a general stochastic model for gene regulatory networks, which can describe bursty gene expression as observed in individual cells.
Instead of using ordinary differential equations, for which cells would structurally all behave the same way, we adopt a more detailed point of view including stochasticity as a fundamental component through the two-state promoter model.
{This model is but a simplification of the complexity of the real molecular processes~\cite{Coulon2010}. Modifications have been proposed, from the existence of a refractory period~\cite{Suter2011} to its attenuation by nuclear buffering~\cite{Battich2015}. In bacteria, the two states originate from the accumulation of positive supercoiling on DNA which stops transcription~\cite{Chong2014}. In eukaryotes, although its molecular basis is not quite understood, the two-state model is a remarkable compromise between simplicity and the ability to capture real-life data~\cite{Raj2006, Vinuelas2013, Kim2013, Albayrak2016, Richard2016}.}
Our PDMP framework appears to be conceptually very similar to the \emph{random dynamical system} proposed in~\cite{Antoneli2016} but it has two major advantages: time does not have to be discretized, and the mathematical analysis is significantly easier.
We also note {that a similar framework appears in~\cite{Potoyan2015,Hufton2016} and} that a closely related PDMP -- which can be seen as the limit of our model for infinitely short bursts -- has recently been described in~\cite{Pajaro2017}.

\bigskip

We then derive an explicit approximation of the stationary distribution and propose to use it as a statistical likelihood to infer networks from single-cell data.
The main ingredient is the separation of three physical timescales -- chromatin, promoter/RNA, and proteins -- and the core idea is to use the self consistent proteomic field approximation from~\cite{Sasai2003,Walczak2005} in a slightly simpler mathematical framework, providing fully explicit formulas that make possible the massive computations usually needed for parameter inference. From this viewpoint, it is a rather simple approach and we hope it can be adapted or improved in more specific contexts, for example in the study of lineage commitment~\cite{Teles2013}. Besides, the main framework does not necessarily has to include an underlying chromatin model and thus it can in principle also be used to describe gene networks in procaryotes.

\subsection*{Mechanistic modelling and statistical inference}

An important quality of the PDMP network model is that the simulation algorithm is comparable in speed with classic ODE and diffusion systems, while providing an effective approximation of the “perfect”, fully discrete, molecular counterpart~\cite{Crudu2009,Lin2016}. It is worth noticing that the PDMP -- at least the promoter-mRNA system -- naturally appears as an example of Poisson representation~\cite{Schnoerr2016,Dattani2017}, that is, not a simple approximation but rather the core component of the \emph{exact} distribution of the discrete molecular model.
Furthermore, such a simulation speed allowed us to compare our approximate likelihood with the true likelihood for a simple two-gene toggle switch, giving excellent results (Fig.~\ref{Exemple_Hartree_1}). This obviously does not constitute a proof of robustness for every network: a proper quantitative (theoretical or numeric) comparison is beyond the scope of this article but would be extremely valuable. Intuitively, it should work for any number of genes, provided that interactions are not too strong.

\bigskip

Besides, some widely used ODE frameworks~\cite{Marbach2010,Mizeranschi2015,Ocone2015} can be seen as the fast-promoter limit of the PDMP model: this limit may not always hold in practice, especially in the bursty regime. In particular, Fig.~\ref{sample_path_1} highlights the risk of using mRNA levels as a proxy for protein levels. It also explains why ordering single-cell mRNA measurements by pseudo-time may not always be relevant, as found in~\cite{Richard2016}.
In~\cite{Ocone2013}, the authors use a hybrid model of gene expression to infer regulatory networks: it is very close to the diffusion limit of our reduced model~\eqref{eq_network_PDMP_red} with the difference that the discrete component, called “promoter” by the authors, would correspond to the “frequency mode” in the present article, as visible for proteins in Fig.~\ref{sample_path_1}. From such a perspective, our approach adds a description of bursty mRNA dynamics that allows for fitting single-cell data such as in Fig.~\ref{data_fitting}.

\bigskip

Finally, our method performed well for simple two-gene networks (Fig.~\ref{test_networks}), showing that part of the causal information remains present in the stationary distribution: this suggests that it is indeed possible to retrieve network structures with a mechanistic interpretation, even from bursty mRNA data.

\subsection*{Perspectives}

We focused here on presenting the key ideas behind the general network model and the inference method: the logical next step is to apply it to real data and with a larger number of genes, which is the subject of work in progress in our group.
{In particular, we propose a functional preprocessing phase, detailed in Appendix~\ref{appendix_data}, that only requires the knowledge of the ratio $d_{0,i}/d_{1,i}$ to estimate all the relevant parameters before inferring $\theta$. The ratio between protein and mRNA degradation rates (or half-lives) hence appears to be the minimum required for such a mechanistic approach to be relevant. Depending upon the species, mRNA and protein half-lives values can be found in the literature (see e.g.~\cite{Schwanhausser2011} for human proteins half-lives), or should be estimated from ad hoc experiments.

\bigskip

From a computational point of view, the main challenge is the algorithmic complexity induced by the fact that proteins are not observed and have to be treated as latent variables. There is a priori no possibility of reducing this without loosing too much accuracy, and therefore some finely optimized algorithms may be required to make the method scalable.
Furthermore, the identifiability properties of the interaction matrix $\theta$ seem difficult to derive theoretically. In this paper we focused on the stationary distribution for simplicity: importantly, several aspects such as time dependence (computing the Hartree approximation in transitory regime) or perturbations (changing the cell's medium or performing knockouts~\cite{Pinna2010}, which can be naturally embedded in our framework) could greatly improve the practical identifiability.}

\bigskip

From a biological point of view, our model does not really describe individual cells but rather a concatenation of trajectories obtained by following cells throughout divisions. Experiments suggest that it should be a relevant approximation, providing one considers mRNA and proteins levels in terms of concentrations instead of molecule numbers~\cite{Corre2014}, which is made possible by the PDMP framework. In this view, the cell cycle results in increasing the apparent degradation rates -- because of the increase in cell volume followed by division -- and thus plays a crucial role for very stable proteins. However, at such a description level, many aspects of possible compensation mechanisms~\cite{Padovan-Merhar2015} and chromatin dynamics~\cite{Hathaway2012} remain to be elucidated.
Regarding the latter aspect, our abstract chromatin states were not modeled from real-life data -- chromatin composition for instance -- but our approach is relevant in that partitioning into dual-type chromatin states as we did is now known as a pervasive feature of all eukaryotic genomes~\cite{Fourel2004,Kueng2013,Rao2014,Obersriebnig2016}.

\section*{Acknowledgments}

We thank Geneviève Fourel (LBMC/ENSL) for critical reading of the manuscript and Anne-Laure Fougères (ICJ) for her constant support.
We also thank all members of the SBDM and Dracula teams for enlightening discussions, and BioSyL Federation and Ecofect Labex for inspiring scientific events. This work was supported by funding from the Institut Rh\^{o}nalpin des Syst\`{e}mes Complexes (IXXI) and from French agency ANR (ICEBERG; ANR-IABI-3096).

\clearpage


\begin{appendices}

\section{First simplifications}
\label{appendix_timescales}

\subsection{Normalizing the PDMP network model}

In this section we detail the normalization of our network model. Recall that the original model is defined by
\begin{equation}%
\label{supp_eq_network_PDMP}
\left\{\begin{aligned}
E_i(t) &: 0 \xrightarrow[]{\koni} 1, \;\; 1 \xrightarrow[]{\koffi} 0 \\
{M_i}'(t) &= s_{0,i} {E_i(t)} - d_{0,i} {M_i(t)} \\
{P_i}'(t) &= s_{1,i} {M_i(t)} - d_{1,i} {P_i(t)}
\end{aligned}\right.
\end{equation}
where $\koni = \koni(P_1,\dots,P_n)$ and $\koffi = \koffi(P_1,\dots,P_n)$.
First we observe that, given an initial condition
\[(E_1^0,\dots,E_n^0)\in \{0,1\}^n, \quad (M_1^0,\dots,M_n^0) \in \prod_{i=1}^n \left[0,\frac{s_{0,i}}{d_{0,i}}\right], \quad (P_1^0,\dots,P_n^0) \in \prod_{i=1}^n \left[0,\frac{s_{0,i}s_{1,i}}{d_{0,i}d_{1,i}}\right] ,\]
the system stays in this set for all $t > 0$, and we introduce the dimensionless variables:
\[\overline{M}_i = \frac{d_{0,i}}{s_{0,i}} M_i \in [0,1] \quad \text{and} \quad \overline{P}_i = \frac{d_{0,i}d_{1,i}}{s_{0,i}s_{1,i}} P_i \in [0,1] .\]
Then, since $s_{0,i}$, $s_{1,i}$, $d_{0,i}$ and $d_{1,i}$ are constants, we get
\[{\overline{M}_i}'(t) = \frac{d_{0,i}}{s_{0,i}} {M_i}'(t) = d_{0,i}\left({E_i(t)} - \frac{d_{0,i}}{s_{0,i}} M_i(t)\right) = d_{0,i}\left({E_i(t)} - \overline{M}_i(t)\right)\]
and
\[{\overline{P}_i}'(t) = \frac{d_{0,i}d_{1,i}}{s_{0,i}s_{1,i}} {P_i}'(t) = d_{1,i} \left(\frac{d_{0,i}}{s_{0,i}} {M_i(t)} - \frac{d_{0,i}d_{1,i}}{s_{0,i}s_{1,i}} {P_i(t)}\right) = d_{1,i}\left(\overline{M}_i(t) - \overline{P}_i(t)\right) .\]
As a result, we obtain the normalized model:
\begin{equation}%
\label{supp_eq_network_PDMP_norm}
\left\{\begin{aligned}
E_i(t) &: 0 \xrightarrow[]{\overline{\koni}} 1, \;\; 1 \xrightarrow[]{\overline{\koffi}} 0 \\
{\overline{M}_i}'(t) &= d_{0,i} \left({E_i(t)} - {\overline{M}_i(t)}\right) \\
{\overline{P}_i}'(t) &= d_{1,i} \left({\overline{M}_i(t)} - {\overline{P}_i(t)}\right)
\end{aligned}\right.
\end{equation}
where the rescaled interaction function $\overline{\koni}$ is defined by
\[\overline{\koni}(\overline{P}_1,\dots,\overline{P}_n) = \koni\left(\frac{s_{0,1}s_{1,1}}{d_{0,1}d_{1,1}}\overline{P}_1,\dots,\frac{s_{0,n}s_{1,n}}{d_{0,n}d_{1,n}}\overline{P}_n\right)\]
and $\overline{\koffi}$ is defined analogously.
It is straightforward to see that, given a path
\[(E_i(t),\overline{M}_i(t),\overline{P}_i(t))_i\]
of the normalized model~\eqref{supp_eq_network_PDMP_norm}, the corresponding path of the original model~\eqref{supp_eq_network_PDMP} is 
\[\left(E_i(t),\frac{s_{0,1}}{d_{0,1}}\overline{M}_i(t),\frac{s_{0,1}s_{1,1}}{d_{0,1}d_{1,1}}\overline{P}_i(t)\right)_i .\]
In this sense, both models are equivalent: in the main text and in the next sections, we always consider model~\eqref{supp_eq_network_PDMP_norm} but forget the “bars” to keep the notations simple.

\subsection{Separating mRNA and protein timescales}

Here we justify the reduced network model involving only promoters and proteins, which is valid when $d_{1,i} \ll d_{0,i}$ for all gene $i$. A full proof is beyond the scope of this article but we provide a heuristic explanation.
We temporarily drop the $i$ index for simplicity.
Let $t_1 \geqslant t_0 \geqslant 0$ and $E\in\{0,1\}$, and suppose $E(t) = E$ for all $t\in[t_0,t_1]$.
Moreover, let $M_0 = M(t_0) \in [0,1]$ and $P_0 = P(t_0)\in [0,1]$.
If $d_1 < d_0$, the solution of the linear ODE system
\begin{equation*}%
\left\{\begin{aligned}
M' &= d_0(E - M) \\
P' &= d_1(M - P)
\end{aligned}\right.
\end{equation*}
is given for $t\in[t_0,t_1]$ by
\begin{equation*}%
\left\{\begin{aligned}
M(t) &= E + (M_0 - E)e^{-d_0(t-t_0)} \\
P(t) &= E + (P_0 - E)e^{-d_1(t-t_0)} + \frac{d_1}{d_0-d_1}(M_0 - E) \left(e^{-d_1(t-t_0)} - e^{-d_0(t-t_0)}\right)
\end{aligned}\right.
\end{equation*}
Hence, if $d_1 \ll d_0$, we have
\[P(t) \approx E + (P_0 - E)e^{-d_1(t-t_0)}\]
using the fact that $|M_0-E|\leqslant 1$ and $|e^{-d_1(t-t_0)} - e^{-d_0(t-t_0)}|\leqslant 1$, and thus $P(t)$ approximates the solution of the differential equation ${P}' = d_{1} \left({E} - {P}\right)$.

\section{Hartree approximation}
\label{appendix_hartree}

\subsection{Hartree approximation for the PDMP model}

Before deriving the approximation, we introduce some notation.
Let $n$ be the number of genes in the network, $\Ee = \{0,1\}^n$ and $\Omega = {(0,1)}^n$. At time $t$, promoter and protein configurations are denoted by $E_t = (e_1,\dots,e_n) = e \in \Ee$ and $P_t = (y_1,\dots,y_n) = y \in \Omega$, respectively.
The distribution of $(E_t, P_t)$ then evolves along time according to its Kolmogorov forward (aka master) equation, which is a linear partial differential equation (PDE) system in our case. This system is high dimensional ($|\Ee|=2^n$, the number of possible promoter configurations) but the associated linear operator contains lots of zeros. Using the tensor product notation $\otimes$, one can write down the equation in a compact form:
\begin{equation}%
\label{pde}
\frac{\dr u}{\dr t} \; + \; \sum_{i=1}^n \frac{\dr \left(F_{i} u\right)}{\dr y_i} \; = \; \sum_{i=1}^n K_i u
\end{equation}
where $u(t,y) = (u_e(t,y))_{e\in\Ee} \in \R^{2^n} \simeq (\R^2)^{\otimes n}$ represents the probability density function (pdf) of $(E_t,P_t)$, and matrices $F_{i}(y_i)$, $K_i(y)\in\Mm_{2^n}(\R) \simeq \Mm_{2}(\R)^{\otimes n}$ are defined by
\[F_{i}(y_i) = I_2 \otimes \cdots \otimes \underbrace{F^{(i)}(y_i)}_{i} \otimes \cdots \otimes I_2 ,\qquad K_i(y) = I_2 \otimes \dots \otimes \underbrace{K^{(i)}(y)}_{i} \otimes \dots \otimes I_2 \]
with
\[F^{(i)}(y_i) = \left(\begin{array}{cc}-d_{1,i}y_i & 0 \\0 & d_{1,i}(1-y_i)\end{array}\right) \quad\text{et} \quad K^{(i)}(y) = \left(\begin{array}{cc}-\koni(y) & \koffi(y) \\\koni(y) & -\koffi(y)\end{array}\right) .\vspace{2mm}\]
The sum in the left side of equation~\eqref{pde} clearly corresponds to a deterministic transport term, while the right side corresponds to the stochastic transitions between promoter configurations.

Furthermore, the PDE system comes with the boundary condition
\begin{equation}%
\label{cond_boundary}
\forall i\in\{1,\dots,n\},\quad F_i u = 0 \quad \text{on } \dr \Omega
\end{equation}
and the probability condition
\begin{equation}%
\label{cond_density}
u \geqslant 0 \quad \text{and}\quad \forall t\in\R_+,\quad \sum_{e\in\Ee}\int_\Omega u_e(t,y)\intd{y} = 1.
\end{equation}

The self-consistent “Hartree” approximation consists in splitting this $2^n$-dimensional problem into $n$ independent $2$-dimensional problems by “freezing” the $y_j$ for $j\neq i$ where $i$ is fixed, and then gathering the solutions by taking their tensor product to produce an approximation of the true pdf (see~\cite{Walczak2005} for a heuristic explanation in the discrete protein setting).
More precisely, one reduced problem is derived for each gene $i$ from~\eqref{pde}-\eqref{cond_boundary}-\eqref{cond_density}:
\begin{equation}%
\label{pde_hartree}
\frac{\dr u^i}{\dr t} + \frac{\dr (F^{(i)}u^i)}{\dr y_i} = K^{(i)}u^i
\end{equation}
where $u^i(t,y) = (u_0^i(t,y),u_1^i(t,y))^\top \in\R_+^2$ satisfies the initial condition $u^i(0,y) = u^{i,0}(y)$, the boundary condition $F^{(i)}(y_i)u^i(y) \to 0$ when $y_i \to 0$ or $1$, and the probability condition $\int_0^1 [u_0^i(t,y) + u_1^i(t,y)]\intd{y_i} = 1$ for all $t\geqslant 0$ and $y_1,\dots,y_{i-1},y_{i+1},\dots,y_n\in (0,1)$.
Therefore, each $u^i$ is a pdf with respect to $(e_i,y_i) \in \{0,1\}\times (0,1)$ but not on $\Ee\times\Omega$. Finally, the Hartree approximation is given by
\begin{equation}%
\label{sol_hartree}
u(t,y) \approx \bigotimes_{i=1}^n u^i(t,y)
\end{equation}
where the equality holds if for all $i$, $\koni$ and $\koffi$ only depend on $y_i$.

\subsection{Solving the reduced problem}

For the moment, the time-dependent closed-form solution of~\eqref{pde_hartree} is unavailable, but the unique stationary solution can be easily obtained if one knows a primitive of
\[\lambda_i : y_i \mapsto \frac{\koni(y)}{d_{1,i}y_i} - \frac{\koffi(y)}{d_{1,i}(1-y_i)}\]
which is the nonzero eigenvalue of the matrix $M^{(i)} = K^{(i)}(F^{(i)})^{-1}$. Indeed, letting $v^i = F^{(i)}u^i$, the stationary equation for $v^i$ from~\eqref{pde_hartree} becomes
\[\frac{\dr v^i}{\dr y_i} = M^{(i)}v^i\]
and then, crucially using the fact that $M^{(i)}$ has a constant eigenvector $(-1,1)^\top$ associated with eigenvalue $\lambda_i$ (the other eigenvalue being 0), one can check that $v^i = e^{\varphi_i} (-1,1)^\top$ is a solution when $\frac{\dr \varphi_i}{\dr y_i} = \lambda_i$. If one has such a $\varphi_i$, the stationary solution of~\eqref{pde_hartree} is given by
\begin{equation}%
\label{sol_hartreei_cst}
u_0^i(y) = {Z_i}^{-1} y_i^{-1} \exp(\varphi_i(y)) \quad \text{and} \quad u_1^i(y) = {Z_i}^{-1} (1-y_i)^{-1} \exp(\varphi_i(y))
\end{equation}
where ${Z_i}$ is the normalizing constant (which may still depend on $y_j$ for $j\neq i$).
Note that the existence of a positive constant $\alpha$ such that $\min(\koni, \koffi)\geqslant\alpha$ imposes the limit 0 for $\exp(\varphi_i(y))$ when $y_i\to 0$ or $1$, and thus the boundary condition is satisfied.
We also obtain the promoter probabilities $p_{0,i} = p(e_i = 0) = Z_{0,i}/{Z_i}$ and $p_{1,i} = p(e_i = 1) = Z_{1,i}/Z_i$ where $Z_{0,i} = \int_0^1 y_i^{-1} \exp(\varphi_i(y)) \intd{y_i}$, $Z_{1,i} = \int_0^1 (1-y_i)^{-1} \exp(\varphi_i(y)) \intd{y_i}$ and $Z_i = Z_{0,i}+Z_{1,i}$.

\bigskip

In particular, when $\koni$ and $\koffi$ do not depend on $y_i$ (i.e. no self-interaction), we get
\[\varphi_i(y) = \frac{\koni(y)}{d_{1,i}}\log(y_i) + \frac{\koffi(y)}{d_{1,i}}\log(1-y_i)\]
which gives the classical solution
\begin{equation}%
\label{sol_hartreei_classic}
u_{0}^i(y) = \frac{b_i}{a_i+b_i} \cdot \frac{y_i^{a_i-1} (1-y_i)^{b_i}}{\Beta(a_i,b_i+1)}\quad \text{and} \quad u_{1}^i(y) = \frac{a_i}{a_i+b_i} \cdot \frac{y_i^{a_i} (1-y_i)^{b_i-1}}{\Beta(a_i+1,b_i)}
\end{equation}
with $a_i = {\koni(y)}/{d_{1,i}}$ and $b_i = {\koffi(y)}/{d_{1,i}}$. This form makes clear the promoter probabilities $p_{0,i}$ and $p_{1,i}$ and the conditional distributions of protein $y_i$ given the promoter state $e_i=0$ or $1$, both being Beta distributions. Since the state is usually not observed, one usually considers the marginal pdf of $y_i$, which is also a Beta:
\begin{equation}%
\label{sol_hartreei_classic_marg}
\underline{u}^i(y) = u_{0}^i(y) + u_{1}^i(y) = \frac{y_i^{a_i-1} (1-y_i)^{b_i-1}}{\Beta(a_i,b_i)}.
\end{equation}
Note that the conditional distribution of mRNA given proteins also has the form~\eqref{sol_hartreei_classic_marg} since the PDMP equation is the same, although the argument is not the Hartree approximation but rather the more common quasi-steady state assumption.

\subsection{Protein marginal distribution}

Given the form of the solution~\eqref{sol_hartreei_cst}, it is in fact always straightforward to integrate over promoters, even for the full (stationary) Hartree approximation~\eqref{sol_hartree}, and we finally obtain
\begin{equation}%
\label{sol_hartree_marg}
\underline{u}(y) = \sum_{e\in\Ee}u_e(y) \approx \sum_{e\in\Ee}\left[\bigotimes_{i=1}^n u^i(y)\right]_e = \sum_{e\in\Ee}\prod_{i=1}^n \frac{\exp(\varphi_i(y))}{Z_i(y)|e_i-y_i|} = \prod_{i=1}^n \frac{\exp(\varphi_i(y))}{Z_i(y)y_i(1-y_i)}
\end{equation}
where we recalled the possible dependence of $Z_i$ on some $y_j$.
Hence, when $\varphi_i$ and $Z_i$ are known functions, one gets a fully explicit approximation of the joint protein distribution.

\section{Explicit interactions}
\label{appendix_interactions}

Here we derive an explicit form for the interactions between genes, starting from a coarse-grained biochemical model.
That is, for a given gene $i$, we focus on defining functions $\koni(y_1,\dots,y_n)$ and $\koffi(y_1,\dots,y_n)$ where $y_1,\dots,y_n$ denote the protein quantities. For simplicity, we drop the $i$ index in this section when there is no ambiguity.

\subsection{Simple biochemical model}

The basic idea is to slightly refine the two-state model of gene expression: in addition to the usual switching reactions (whose rates are $\kon$ and $\koff$), we consider a set of reversible transitions between some chromatin states (e.g. describing enhancer regions). Each chromatin state is then associated with a particular rate for the promoter activation reaction. For simplicity, we consider only two cases: a high rate $k_1$ (the chromatin will be said \emph{permissive}) and a low rate $k_0 \ll k_1$ (the chromatin will be said \emph{non-permissive}). Once active, the promoter can switch off at a rate that is supposed to be independent from chromatin states.
Finally, we assume that the chromatin transitions are due to fast interactions with ambiant proteins (binding, hit-and-run, etc.) so that the promoter-switching reactions always see chromatin in its quasi-stationary state. Effective rates $\kon$ and $\koff$ can therefore be obtained by averaging over chromatin states: this way, $\koff$ is still a constant and $\kon$ is now defined by
\[\kon = k_0p_0 + k_1p_1\]
where $p_0$ (resp. $p_1$) is the probability of the chromatin being non-permissive (resp. permissive). 

\bigskip

We now define an explicit model for chromatin dynamics and compute its stationary distribution to derive $p_0$ and $p_1$ as functions of $y_1,\dots,y_n$.
We consider $2^n$ permissive configurations and $2^n$ non-permissive configurations as follows: for all $I\subset \Gg$ where $\Gg = \{1,\dots,n\}$, species $C_I$ (resp. $C^*_I$) stands for the chromatin being non-permissive (resp. permissive) and in state $I$. The underlying physics are the following: the chromatin has two “basal” configurations $C_{\emptyset}$ (non-permissive) and $C^*_{\emptyset}$ (permissive), which describe dynamics when no protein is present, according to the reactions
\[C_{\emptyset} \xrightarrow[]{\alpha} C^*_{\emptyset} , \quad C^*_{\emptyset} \xrightarrow[]{\beta} C_{\emptyset} .\]
Then, each protein $P_j$ is able to modify the chromatin state through a “hit-and-run” reaction, which is kept in memory by encoding the index $j$ in the list $I$, giving the state $C_I$ or $C^*_I$.
Eventually, this memory can be lost by “emptying” $I$ step by step (going back to the basal configuration).
That is, for all $I\subset\Gg$ and $j\in \Gg \setminus I$, we consider the reactions
\[C^*_I + P_j \xrightarrow[]{a_j} C^*_{I\cup j} + P_j , \quad C^*_{I\cup j} \xrightarrow[]{b_j} C^*_I ,\]
\[C_I + P_j \xrightarrow[]{c_j} C_{I\cup j} + P_j , \quad C_{I\cup j} \xrightarrow[]{d_j} C_I .\]

\bigskip

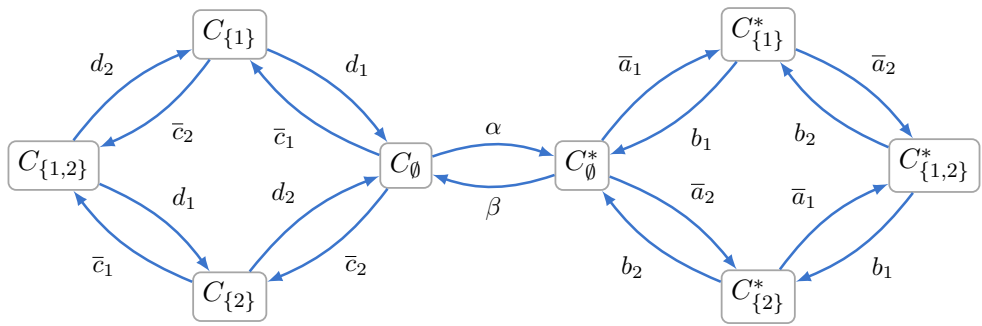
\begin{figure}[htbp]
\begin{center}
\resizebox{0.8\textwidth}{!}{\begin{tikzpicture}[scale=1.2,line width = 1]
\tikzstyle{noeud}=[draw, rounded corners=3pt, line width = 0.7, color = gray!70, text = black]
\tikzstyle{ligne}=[{->}, >=latex, color = bleuclair]
\node[noeud] (000) at (-1,0) {$C_{\emptyset}$};
\node[noeud] (001) at (-3,-1.5) {$C_{\{2\}}$};
\node[noeud] (010) at (-3,1.5) {$C_{\{1\}}$};
\node[noeud] (011) at (-5,0) {$C_{\{1,2\}}$};
\node[noeud] (100) at (1,0) {$C_{\emptyset}^*$};
\node[noeud] (101) at (3,-1.5) {$C^*_{\{2\}}$};
\node[noeud] (110) at (3,1.5) {$C^*_{\{1\}}$};
\node[noeud] (111) at (5,0) {$C^*_{\{1,2\}}$};
\draw[ligne] (000) to[bend left = 19] node[midway, color=black, above]{\small $\alpha$} (100);
\draw[ligne] (100) to[bend left = 19] node[midway, color=black, below]{\small $\beta$} (000);

\draw[ligne] (000) to[bend left = 15] node[midway, color=black, below right]{\small $\overline{c}_2$} (001);
\draw[ligne] (001) to[bend left = 15] node[midway, color=black, above left]{\small $d_2$} (000);
\draw[ligne] (000) to[bend left = 15] node[midway, color=black, below left]{\small $\overline{c}_1$} (010);
\draw[ligne] (010) to[bend left = 15] node[midway, color=black, above right]{\small $d_1$} (000);
\draw[ligne] (001) to[bend left = 15] node[midway, color=black, below left]{\small $\overline{c}_1$} (011);
\draw[ligne] (011) to[bend left = 15] node[midway, color=black, above right]{\small $d_1$} (001);
\draw[ligne] (010) to[bend left = 15] node[midway, color=black, below right]{\small $\overline{c}_2$} (011);
\draw[ligne] (011) to[bend left = 15] node[midway, color=black, above left]{\small $d_2$} (010);

\draw[ligne] (100) to[bend left = 15] node[midway, color=black, above right]{\small $\overline{a}_2$} (101);
\draw[ligne] (101) to[bend left = 15] node[midway, color=black, below left]{\small $b_2$} (100);
\draw[ligne] (100) to[bend left = 15] node[midway, color=black, above left]{\small $\overline{a}_1$} (110);
\draw[ligne] (110) to[bend left = 15] node[midway, color=black, below right]{\small $b_1$} (100);
\draw[ligne] (101) to[bend left = 15] node[midway, color=black, above left]{\small $\overline{a}_1$} (111);
\draw[ligne] (111) to[bend left = 15] node[midway, color=black, below right]{\small $b_1$} (101);
\draw[ligne] (110) to[bend left = 15] node[midway, color=black, above right]{\small $\overline{a}_2$} (111);
\draw[ligne] (111) to[bend left = 15] node[midway, color=black, below left]{\small $b_2$} (110);
\end{tikzpicture}}
\caption{Chromatin states and transitions rates in the case of $n=2$ proteins.}
\label{fig_chromatin}
\end{center}
\end{figure}

The system then evolves with $[C_I], [C_I^*] \in \{0,1\}$ and $\sum_I [C_I] + [C_I^*] = 1$, so that only one molecule is present at a time: its species therefore entirely describes the state of the system. Mathematically, we obtain a standard jump Markov process with $2^{n+1}$ states.
For example, the case $n=2$ leads to the scheme of Figure~\ref{fig_chromatin}, writing $\overline{a}_j = a_j [P_j]$ and $\overline{c}_j = c_j [P_j]$ for simplicity.
The underlying idea is that, depending on $a_j$, $b_j$, $c_j$ and $d_j$, proteins will tend to {stabilize} the chromatin either in a permissive configuration or in a non-permissive one -- providing notions of \emph{activation} and \emph{inhibition}.
The basal reactions with rates $\alpha$ and $\beta$ sum up what we do not observe (i.e. what is likely to happen for the chromatin when none of the $P_j$ are present).

\bigskip

\bigskip

\subsection{Stationary distribution}

Letting $\Ss = \{0,1\}^{n+1}$, each state can be coded by a vector $s = (s_0, s_1,\dots, s_n)\in \Ss$ where $s_0 = 1$ if the chromatin is permissive and 0 otherwise, and for $j\geqslant 1$, $s_j = 1$ if it has been modified by protein $P_j$ and 0 otherwise.
If all rates are positive, the system has a unique stationary distribution $\pi$ which can be exactly computed from the master equation. More precisely, the probability $\pi_s$ of the chromatin being in state $s\in\Ss$ is given by
\[\pi_s =
\begin{cases}
Z^{-1} \alpha \prod_{j=1}^n (\lambda_j [P_j] s_j + 1-s_j) & \text{if } s_0 = 1\\
Z^{-1} \beta \prod_{j=1}^n (\mu_j [P_j] s_j + 1-s_j) & \text{if } s_0 = 0
\end{cases}\]
where $\lambda_j = a_j/b_j$, $\mu_j = c_j/d_j$ and $Z$ is a normalizing constant.
Now going back to our initial intention of computing $\kon$, we are only interested in the probability for the chromatin to be permissive,
\[p_1 = \sum_{s_1,\dots, s_n} \! \pi_{(1,s_1,\dots, s_n)} = Z^{-1} \alpha \sum_{s_1,\dots, s_n} \prod_{j=1}^n (\lambda_j [P_j] s_j + 1-s_j) ,\]
and the probability for the chromatin to be non-permissive,
\[p_0 = \sum_{s_1,\dots, s_n} \! \pi_{(0,s_1,\dots, s_n)} = Z^{-1} \beta \sum_{s_1,\dots, s_n} \prod_{j=1}^n (\mu_j [P_j] s_j + 1-s_j) .\]
Observing that each product term only depends on one $s_j$, these formulas collapse to
\[p_1 = Z^{-1} \alpha \prod_{j=1}^n (\lambda_j [P_j] +1) , \quad p_0 = Z^{-1} \beta \prod_{j=1}^n (\mu_j [P_j] + 1)\]
and the distribution condition $p_0 + p_1 = 1$ gives $Z = \alpha \prod_{j=1}^n (\lambda_j [P_j] +1) + \beta \prod_{j=1}^n (\mu_j [P_j] +1)$.
We finally get
\begin{equation}%
\label{eq_kon_base}
\kon = \frac{k_0 \beta \prod_{j=1}^n (\mu_j [P_j] +1) + k_1\alpha \prod_{j=1}^n (\lambda_j [P_j] +1)}{\beta \prod_{j=1}^n (\mu_j [P_j] +1) + \alpha \prod_{j=1}^n (\lambda_j [P_j] +1)} .
\end{equation}


From this formula, it is straightforward to see that $\kon$ will actually depend on a protein $P_j$ only if $\lambda_j \neq \mu_j$, that is, when reactions involving $P_j$ have unbalanced speeds and tend to favor either permissive configurations ($\lambda_j > \mu_j$) or non-permissive configurations ($\lambda_j < \mu_j$).

\subsection{Higher order interactions}

So far we only considered that the $P_j$ were interacting as monomers. If they in fact interact after forming dimers or other complexes, and if such complex-forming reactions are even faster than chromatin dynamics, one can take this into account by simply replacing $[P_j]$ in equation~\eqref{eq_kon_base} with a function of $[P_j]$ corresponding to the quasi-stationary concentration of the complex.
This approximation seems to be relevant to capture the overall dependence of $\kon$ on the proteins, the main point being to use a continuous description (e.g. rate equations) for proteins, which are abundant, while keeping a discrete (stochastic) description for chromatin.
We chose to replace $[P_j]$ with $[P_j]^{m_j}$ where $m_j > 0$, which gives our model a general Hill-type form. Note that $m_j = 2$ (resp. $m_j = 3$) may represent a correct approximation for $P_j$ interacting as a dimer (resp. a trimer) but in general $m_j$ does not necessarily have to be an integer.

\subsection{The case of auto-activation}

A this stage, it is possible to implement self-interaction for gene $i$ by taking $\lambda_i \neq \mu_i$ in~\eqref{eq_kon_base} but this leads to obvious identifiability issues: in stationary state, one cannot really distinguish between auto-activation, auto-inhibition and basal level. To cope with these, we restrict ourselves to auto-activation by setting $c_i = d_i = 0$ and keeping only the relevant chromatin states ($C_I^*$ for all $I$, and $C_I$ for $I$ such that $i\notin I$). The system still has a unique stationary distribution and the formula for $\kon$ corresponds to the case $\mu_i = 0$ in~\eqref{eq_kon_base}. Then, starting from the fact that auto-activation is only relevant when the basal level is small enough (for a bistable behaviour to be possible), we take the limit $\alpha \ll 1$ while keeping $\alpha \lambda_i$ fixed: the formula becomes
\begin{equation}%
\label{eq_kon_auto}
\kon = \frac{k_0 \beta \prod_{j\neq i} (\mu_j [P_j]^{m_j} +1) + k_1\alpha \lambda_i [P_i]^{m_i} \prod_{j\neq i} (\lambda_j [P_j]^{m_j} +1)}{\beta \prod_{j\neq i} (\mu_j [P_j]^{m_j} +1) + \alpha \lambda_i [P_i]^{m_i} \prod_{j\neq i} (\lambda_j [P_j]^{m_j}+1)}
\end{equation}
where $m_i > 0$ if gene $i$ activates itself and $m_i = 0$ otherwise.

\subsection{Parameterization for inference}

Parameters of equation~\eqref{eq_kon_auto} are still clearly not identifiable: in order to get a more minimal form, we introduce the following parameterization: $s_j = {\mu_j}^{-1/m_j}$, $\theta_j = \log(\lambda_j/\mu_j)$ for all $j\neq i$, and $s_i = (\beta/\alpha)^{1/m_i}$, $\theta_i = \log(\lambda_i)$. After simplifying~\eqref{eq_kon_auto}, we obtain
\[\kon = \frac{k_0 + k_1 \Phi ([P_i]/s_i)^{m_i}}{1 + \Phi ([P_i]/s_i)^{m_i}}\]
where
\[\Phi = \exp(\theta_i) \prod_{j\neq i} \frac{1 + \exp(\theta_j) ([P_j]/s_j)^{m_j}}{1 + ([P_j]/s_j)^{m_j}} .\]

The new parameters have an intuitive meaning: $s_j$ can be seen as a threshold for the influence by protein $j$, and $\theta_j$ characterizes this influence via its sign and absolute value ($\theta_j = 0$ implying that $\kon$ does not depend on protein $j$), with the exception that $s_i$ and $\theta_i$ aggregate a basal behaviour and an auto-activation strength.

\bigskip

Finally, we recall the notation $y_j = [P_j]$ and reintroduce the index $i$ of the gene of interest and add it to each parameter. Hence, for every gene $i$, the function $\koni$ is defined by:
\begin{equation}
\label{eq_kon_full_s}
\koni(y) = \frac{k_{0,i} + k_{1,i} \Phi_i(y)(y_i/s_{i,i})^{m_{i,i}}}{1 + \Phi_i(y)(y_i/s_{i,i})^{m_{i,i}}}
\end{equation}
with
\begin{equation}
\label{eq_phi_full_s}
\Phi_i(y) = \exp(\theta_{i,i}) \prod_{j\neq i} \frac{1 + \exp(\theta_{i,j}) (y_j/s_{i,j})^{m_{i,j}}}{1 + (y_j/s_{i,j})^{m_{i,j}}} .
\end{equation}

In our statistical framework, we assume that parameters $k_{0,i}$, $k_{1,i}$, $m_{i,j}$ and $s_{i,j}$ are known and we focus on inferring the matrix $\theta = (\theta_{i,j}) \in \Mm_n(\R)$, which is similar to the interaction matrix in usual gene network inference methods.

\subsection{Explicit distribution for an auto-activation model}
\label{sec_autoactiv_model}

Here we derive the stationary distribution for a self-activating gene. For simplicity, we drop the $i$ index. In this model, $\koff$ is constant and we assume that there are some constants $\Phi\geqslant 0$, $m\geqslant 0$, $s>0$ and $k_1\gg k_0 > 0$ such that $\kon$ has the form
\begin{equation*}%
\kon(y) = \frac{k_{0} + k_{1} \Phi (y/s)^{m}}{1 + \Phi (y/s)^{m}}
\end{equation*}
so the stationary distribution can directly be used in the Hartree approximation of the network model~\eqref{eq_kon_full_s}, recalling that $\Phi$ has to be independent of the gene's own protein but can depend on others. Letting $c = (k_{1} - k_{0})/(m d_{1}) > 0$, we are in the case of the explicit solution~\eqref{sol_hartreei_cst} with
\[\varphi(y) = c\log\left(y^{\frac{k_0}{d_1 c}} + \frac{\Phi}{s^m} y^{\frac{k_1}{d_1 c}}\right) + \frac{\koff}{d_{1}}\log(1-y)\]
so the protein distribution is
\begin{equation}%
\label{eq_autoactivation_distrib_0}
\underline{u}(y) = Z^{-1} y^{-1} \left(y^{\frac{k_0}{d_1 c}} + \frac{\Phi}{s^m} y^{\frac{k_1}{d_1 c}}\right)^c (1-y)^{\frac{\koff}{d_{1}}-1} .
\end{equation}

To get a fully explicit result, i.e. to compute $Z$, we shall assume that $c$ is a positive integer. If it is not, one can get a satisfying approximation by taking $c = \lceil(k_{1} - k_{0})/(m d_{1})\rceil$. Then, expanding~\eqref{eq_autoactivation_distrib_0} using the binomial theorem, we obtain
\[Z = \sum_{r = 0}^{c} \binom{c}{r} \Beta(a_{r},b) (\Phi/s^{m})^{r}\]
where $a_{r} = ((c - r)k_{0} + rk_{1})/(d_{1}c)$ and $b = \koff/d_{1}$, and a probabilistic representation of $\underline{u}$ in terms of a mixture of Beta distributions:
\begin{equation}%
\label{eq_autoactivation_distrib}
\underline{u}(y) = \sum_{r=0}^{c} p_{r} f_r(y)
\end{equation}
where $f_r(y) = {y}^{a_{r}-1} {(1-y)}^{b-1}/\Beta(a_{r},b)$ and $p_r = \binom{c}{r} \Beta(a_{r},b) (\Phi/s^{m})^r/Z$.

\bigskip

\begin{figure}[htbp]
\begin{center}
\includegraphics[width=\textwidth]{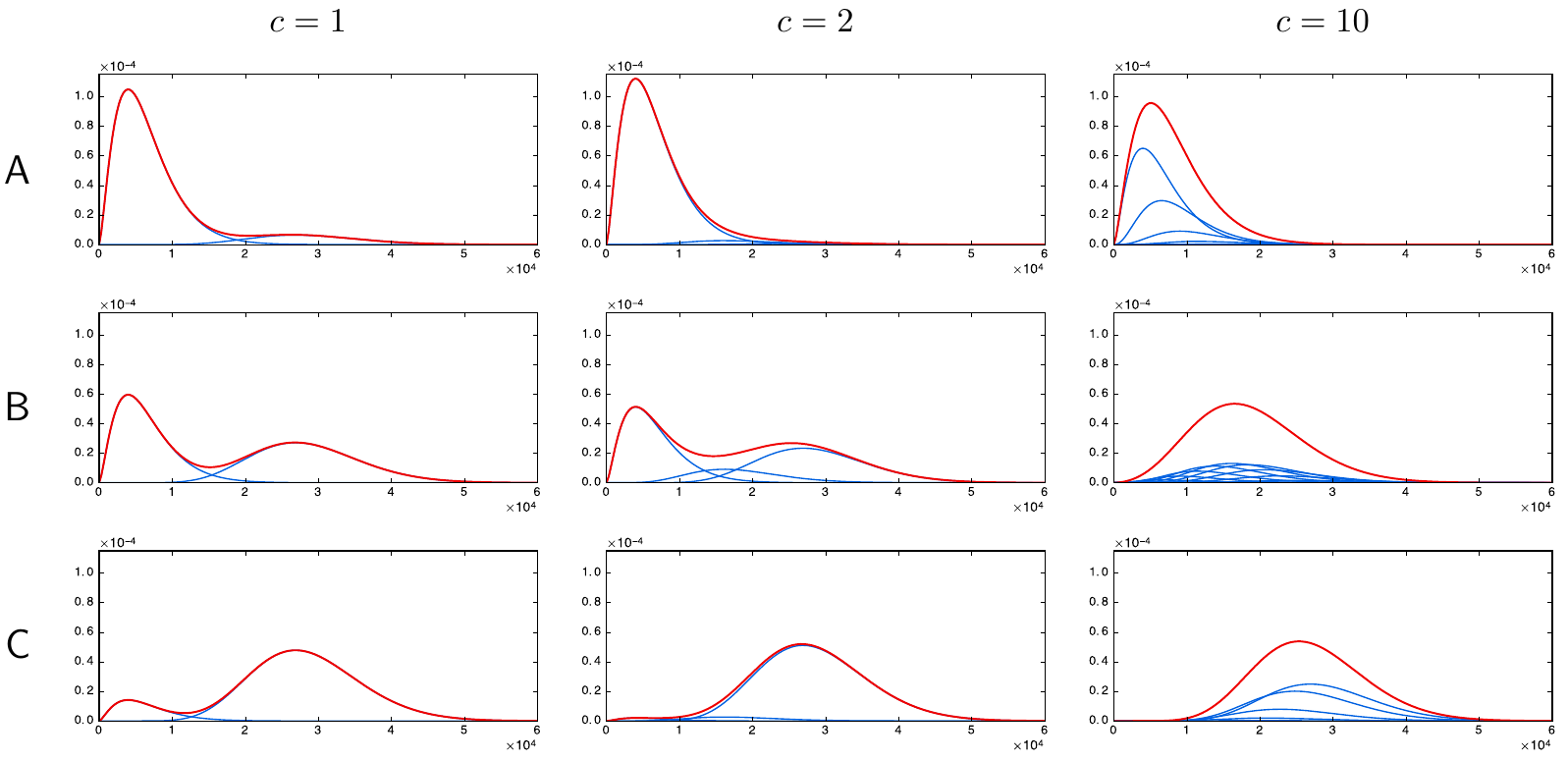}
\caption{Protein stationary distributions (red curves) from the auto-activation model for different values of the input $\Phi$, with $s$ set as in \eqref{eq_s_sym}. The blues curves indicate the underlying weighted Beta distributions in each mixture.
(A) $\Phi = \exp(-2)$, (B) $\Phi = \exp(0) = 1$, (C) $\Phi = \exp(2)$.
The distribution tends to be strongly bimodal for small $c$ values, while large values make the distribution close to the unimodal no-feedback case (constant~$\kon$).
Parameters are $k_0 = 0.25$, $k_1 = 1.25$, $\koff = 7.5$, $d_1 = 0.1$ and, only used for scaling, $s_1 = 10$, $d_0 = 0.5$, $s_0 = 10^3$.}
\label{example_beta_mixtures}
\end{center}
\end{figure}

The dissociation constant $s$ is clearly redundant with the input $\Phi$. We fix the particular value
\begin{equation}%
\label{eq_s_sym}
s = \left(\frac{\Beta(a_c,b)}{\Beta(a_0,b)}\right)^{\frac{1}{mc}} = \left(\frac{\Beta(k_1/d_1,\koff/d_1)}{\Beta(k_0/d_1,\koff/d_1)}\right)^{\frac{d_1}{k_1-k_0}}
\end{equation}
for which the arbitrary neutral case $\Phi = 1$ is “symmetric”, i.e. $p_0 = p_c$.
Note that $s$ actually only depends on the fundamental parameters $k_0$, $k_1$, $\koff$ and $d_1$ (and not on $c$ nor $m$).
Figure~\ref{example_beta_mixtures} shows some examples of the resulting distribution, which can be bimodal or not, depending on the value of $c$ (or equivalently, $m$) when all other parameters are fixed.

\section{EM algorithm for network inference}
\label{appendix_algorithm}

\subsection{EM algorithm for MAP estimation}
\label{sec_EM_MAP}

Here we briefly recall the formulation of the Expectation-Maximization (EM) algorithm for \emph{maximum a posteriori} (MAP) estimation. Consider the probabilistic hierarchical model defined by the distribution of proteins $p(y|\theta)$, the distribution of mRNA given proteins $p(x|y,\theta)$, and a prior distribution $p(\theta)$ on the parameters. Assuming we only observe $x$, we want to infer $\theta$ by MAP estimation, that is, find a mode -- hopefully the highest -- of the posterior distribution $p(\theta\,|\,x)$, which satisfies by Baye's rule:
\[{p(\theta\,|\,x)} = \!\int\!{p(\theta,y\,|\,x)} \intd{y} \quad \text{where} \quad {p(\theta,y\,|\,x)} = p(y\,|\,\theta)p(x\,|\,y,\theta)\frac{p(\theta)}{p(x)} .\]
As $p(\theta\,|\,x)$ has a too complex expression to be efficiently maximized, the EM algorithm rather uses $\ell_\theta(x,y) = \log(p(\theta,y\,|\,x))$ by iteratively computing $\theta^{t+1} = \arg\max_\theta \{Q(\theta,\theta^t)\}$ given $\theta^t$, where
\begin{equation}%
\label{eq_EM_int}
\theta \mapsto Q(\theta,\theta^t) = \int \ell_\theta(x,y) p(y\,|\,x,\theta^t) \intd{y} .
\end{equation}
A well-known result states that at each step we in fact maximize a lower bound of $p(\theta\,|\,x)$, which is the key point of the algorithm and makes it a particular case of “variational method” (see~\cite{Jordan1999} for example). Now, since $p(x)$ (resp. $p(\theta)$) does not depend on $\theta$ (resp. $y$), it turns out that 
\[\arg\max_\theta \{Q(\theta,\theta^t)\} = \arg\max_\theta \{\overline{Q}(\theta,\theta^t) - g(\theta)\}\]
where $g(\theta) = -\log(p(\theta))$ and $\overline{Q}(\theta,\theta^t) = \int [\log p(y\,|\,\theta) + \log p(x\,|\,y,\theta)] p(y\,|\,x,\theta^t) \intd{y}$ is the more standard quantity that appears in the “frequentist” EM algorithm for maximum likelihood estimation. Hence, considering a prior on $\theta$ simply results in adding a penalization term $g(\theta)$ during the M step in the algorithm.

\bigskip

For example, if we assume that $\theta_{i,j}$ for $i\neq j$ are independent and follow Laplace distributions, i.e. $p(\theta) = \prod_{i \neq j} \frac{\lambda}{2} \exp(-\lambda|\theta_{i,j}|)$, then $g(\theta) = \lambda\sum_{i \neq j} |\theta_{i,j}| \;+\; C$ where $C = n(n-1)\log(2/\lambda)$. Since $C$ does not depend on $\theta$, this is equivalent to the standard $L^1$ (lasso) penalization, which is well known to enforce the sparsity of the network.

\subsection{Custom prior on the interactions}
\label{sec_EM_prior}

Here we consider a custom prior to deal with oriented interactions. Indeed, for every pair of nodes $\{i,j\}$ there are two possible interactions with respective parameters $\theta_{i,j}$ and $\theta_{j,i}$, but it is likely that only one is actually present in the true network. Hence, we want $\theta_{i,j}$ and $\theta_{j,i}$ to “compete” against each other so that only one is nonzero after MAP estimation, unless there is enough evidence in the data that both interactions are present.
To this aim, we define the following prior:
\begin{equation}%
\label{eq_prior_network}
p(\theta) \varpropto \exp\left(-\lambda \sum_{i\neq j} |\theta_{i,j}| - \lambda\alpha \sum_{i< j} |\theta_{i,j}\theta_{j,i}| \right)
\end{equation}
with $\lambda,\alpha \geqslant 0$. Thus $\alpha$ can be seen as a competition parameter, the case $\alpha = 0$ leading to the standard lasso penalization parametrized by $\lambda$.

\subsection{The algorithm in practice}
\label{seq_algo_practice}

As visible in~\eqref{eq_EM_int}, the true EM algorithm involves integration against the distribution $p(y\,|\,x,\theta)$, which does not allow for direct numerical integration because of the dimension ($y\in \R^n$). To overcome this problem, a first option is Monte Carlo integration -- typically by MCMC -- leading to a “stochastic EM” algorithm that is slow but accurate if samples are large enough. A faster option consists in approximating $p(y\,|\,x,\theta)$ by its highest mode, i.e. by the Dirac mass $\delta_{\widehat{y}}$ where $\widehat{y} = \arg\max_y\{p(y\,|\,x,\theta)\}$. Then it is worth noticing that since $p(y\,|\,x,\theta) \varpropto p(y\,|\,\theta)p(x\,|\,y,\theta)$, the whole procedure can be seen as performing a coordinate ascent on the function $(\theta,y)\mapsto p(\theta,y\,|\,x)$.
We chose this option for the examples: it is sometimes called “hard” or “classification” EM, since a particular case leads to the well-known $k$-means clustering algorithm~\cite{Celeux1991}.
Unfortunately, theoretical foundations of the true EM algorithm are lost by the hard EM (we do not maximize a lower bound of $p(\theta\,|\,x)$ anymore), but it often gives satisfying results while requiring much less computational time.

\bigskip

In practice, the procedure is the following. Suppose we observe mRNA levels in $m$ independent cells, and let $\xx_k\in\R^n$ (resp. $\yy_k\in\R^n$) denote the mRNA (resp. protein) levels of cell $k$.
In line with sections~\ref{sec_EM_MAP}-\ref{sec_EM_prior} and letting $\xx = (\xx_1,\dots,\xx_m)$ and $\yy = (\yy_1,\dots,\yy_m)$ for simplicity, we define the objective function
\begin{equation}
\label{eq_objective}
\Ff(\yy,\theta) = \ell(\xx,\yy,\theta) - g(\theta)
\end{equation}
where the complete log-likelihood $\ell(\xx,\yy,\theta)$ and the penalization $g(\theta)$ are given by
\[\ell(\xx,\yy,\theta) = \sum_{k=1}^m \log(u(\yy_k,\theta)) + \log(v(\xx_k,\yy_k,\theta)) \quad \text{and} \quad g(\theta) = \lambda \sum_{i\neq j} |\theta_{i,j}| + \lambda\alpha \sum_{i< j} |\theta_{i,j}\theta_{j,i}| ,\]
with $u(y,\theta) = p(y|\theta)$ and $v(x,y,\theta) = p(x|y,\theta)$.

\bigskip

The algorithm then simply consists in iterating the following two steps until convergence:
\begin{align}
\label{eq_algo_hard_EM_E}
\yy^{t+1} &= \arg\max_\yy\{\Ff(\yy,\theta^t)\} \\
\label{eq_algo_hard_EM_M}
\theta^{t+1} &= \arg\max_\theta\{\Ff(\yy^{t+1},\theta)\}
\end{align}
The “approximate E step”~\eqref{eq_algo_hard_EM_E} can be performed using a standard gradient method since $u$ and $v$ are smooth functions of $y$.
The “penalized M step”~\eqref{eq_algo_hard_EM_M} is a non-smooth maximization problem since $g$ is non-smooth, but it can be performed using a proximal gradient method detailed in the next section.
The form of $\ell(\xx,\yy,\theta)$ is such that we just need to compute $\grad\log u$ and $\grad\log v$.

\bigskip

The formulas for $u$ and $v$ derived from the normalized model are given in the main text: they can be applied once the data has been normalized, i.e. after dividing each mRNA $i$ level by $s_{0,i}/d_{0,i}$. In the bursty regime, this scale parameter is neither identifiable nor necessary. Indeed, as explained in the main text, the “Beta-like” distributions collapse to “Gamma-like” ones which we provide below, and for which the scale parameter is identifiable.

\subsubsection{Likelihood form in the basic case}

In the basic case (no self-interaction), we have:
\[\log(u(y,\theta)) = \sum_{i=1}^n (a_i(y,\theta)-1)\log(y_i) - b_i(y,\theta)y_i + a_i(y,\theta)\log(b_i(y,\theta)) -\log\Gamma(a_i(y,\theta))\]
and
\[\log(v(x,y,\theta)) = \sum_{i=1}^n (\widetilde{a}_i(y,\theta)-1)\log(x_i) - \widetilde{b}_i(y,\theta)x_i + \widetilde{a}_i(y,\theta)\log(\widetilde{b}_i(y,\theta)) -\log\Gamma(\widetilde{a}_i(y,\theta))\]
where $a_i = \koni/d_{1,i}$, $b_i = (d_{0,i}/s_{1,i})\times(\koffi/s_{0,i})$, $\widetilde{a}_i = \koni/d_{0,i}$ and $\widetilde{b}_i = \koffi/s_{0,i}$.

\subsubsection{Likelihood form in the auto-activation case}
\label{seq_likelihood_autoactiv_gamma}

In the auto-activation case (section~\ref{sec_autoactiv_model}), the formula for $\log v$ is the same but $\log u$ is given by
\[\log(u(y,\theta)) = \sum_{i=1}^n \log\left(\sum_{r=0}^{c_i} w_{i,r}(y,\theta) {y_i}^{a_{i,r}-1} e^{- b_iy_i} \right) - \log \left(\sum_{r = 0}^{c_i} w_{i,r}(y,\theta) \Gamma(a_{i,r}){b_i}^{-a_{i,r}}\right) \]
where $c_i = \lceil(k_{1,i} - k_{0,i})/(d_{1,i} m_{i,i})\rceil$, $a_{i,r} = ((c_i - r)k_{0,i} + rk_{1,i})/(d_{1,i}c_i)$ and $w_{i,r} = \binom{c_i}{r} (\Phi_i/s_{i,i}^{m_{i,i}})^r$.

\subsubsection{Gradients}

Explicit computation of the gradients is then straightforward (e.g. with $\koffi$ constant and $\koni$, $\Phi_i$ defined by~\eqref{eq_kon_full_s}-\eqref{eq_phi_full_s}) but leads to cumbersome formulas: we implemented them in Scilab and the code is available upon request.

\subsection{Proximal gradient method}
\label{sec_prox_grad}

Here we recall a standard proximal gradient method~\cite{Parikh2013} to solve the M step~\eqref{eq_algo_hard_EM_M} and provide the proximal operator associated with $g(\theta)$. Note that the method seems to converge in practice, even if $g$ is not convex.
It is based on the update
\[\theta^{(k+1)} = \operatorname{prox}_\gamma\left(\theta^{(k)} + \gamma\grad_\theta\ell(\xx,\yy,\theta^{(k)})\right)\]
where $\gamma>0$ is a step size (learing rate) and $\operatorname{prox}_\gamma$ is the proximal operator associated with $g(\theta)$, defined on $\Theta \simeq \R^{n^2-n}$ by
\[\operatorname{prox}_\gamma(\tau) = \arg\min_{\theta \in \Theta} \left\{ g(\theta) + \frac{1}{2\gamma}\sum_{i\neq j}(\theta_{i,j} - \tau_{i,j})^2\right\} .\]
In fact, for any $i,j\in\{1,\dots,n\}$ such that $i\neq j$, one can see that $\theta_{i,j}$ and $\theta_{j,i}$ appear in the minimized quantity as independent of all other $\theta$ components. Hence, one just has to compute
\[\operatorname{prox}_\gamma(\tau_1,\tau_2) = \arg\min_{(\theta_1,\theta_2) \in \R^2} \left\{ \lambda\left(|\theta_1| + |\theta_2| + \alpha|\theta_1\theta_2|\right) + \frac{1}{2\gamma}\left((\theta_1 - \tau_1)^2 + (\theta_2 - \tau_2)^2\right)\right\}\]
and use it for any $(\tau_1,\tau_2) = (\tau_{i,j},\tau_{j,i}) \in\R^2$ to obtain the corresponding components of $\operatorname{prox}_\gamma(\tau)$.
Then, letting $\varepsilon = \lambda\gamma$ and assuming $\gamma$ small enough such that $\alpha\varepsilon < 1$, we obtain
\[\operatorname{prox}_\gamma(\tau_1,\tau_2) = \frac{1}{1-(\alpha \varepsilon)^2} (h_1,h_2)\]
with 9 cases for the value of $(h_1,h_2)$ depending on $(\tau_1,\tau_2)$, given by:
{\small
\begin{enumerate}
\item $\left\{\begin{array}{l}
\tau_1 > \varepsilon \\
\tau_1 > \varepsilon(1 + \alpha(\tau_2 - \varepsilon)) \\
\tau_2 > \varepsilon(1 + \alpha(\tau_1 - \varepsilon))
\end{array}\right.$ \quad $\Rightarrow$ \quad $\left\{\begin{array}{l}
h_1 = \tau_1 - \varepsilon(1 + \alpha(\tau_2 - \varepsilon)) \\
h_2 = \tau_2 - \varepsilon(1 + \alpha(\tau_1 - \varepsilon))
\end{array}\right.$

\item $\left\{\begin{array}{l}
\tau_1 > \varepsilon \\
|\tau_2| \leqslant \varepsilon(1 + \alpha(\tau_1 - \varepsilon))
\end{array}\right.$ \quad $\Rightarrow$ \quad $\left\{\begin{array}{l}
h_1 = \tau_1 - \varepsilon \\
h_2 = 0
\end{array}\right.$

\item $\left\{\begin{array}{l}
\tau_1 > \varepsilon \\
\tau_1 > \varepsilon(1 + \alpha(-\tau_2 - \varepsilon)) \\
\tau_2 < -\varepsilon(1 + \alpha(\tau_1 - \varepsilon))
\end{array}\right.$ \quad $\Rightarrow$ \quad $\left\{\begin{array}{l}
h_1 = \tau_1 - \varepsilon(1 + \alpha(-\tau_2 - \varepsilon)) \\
h_2 = \tau_2 + \varepsilon(1 + \alpha(\tau_1 - \varepsilon))
\end{array}\right.$

\item $\left\{\begin{array}{l}
|\tau_1| \leqslant \varepsilon(1 + \alpha(-\tau_2 - \varepsilon)) \\
\tau_2 < -\varepsilon
\end{array}\right.$ \quad $\Rightarrow$ \quad $\left\{\begin{array}{l}
h_1 = 0 \\
h_2 = \tau_2 + \varepsilon
\end{array}\right.$

\item $\left\{\begin{array}{l}
\tau_1 < -\varepsilon \\
\tau_1 < -\varepsilon(1 + \alpha(-\tau_2 - \varepsilon)) \\
\tau_2 < -\varepsilon(1 + \alpha(-\tau_1 - \varepsilon))
\end{array}\right.$ \quad $\Rightarrow$ \quad $\left\{\begin{array}{l}
h_1 = \tau_1 + \varepsilon(1 + \alpha(-\tau_2 - \varepsilon)) \\
h_2 = \tau_2 + \varepsilon(1 + \alpha(-\tau_1 - \varepsilon))
\end{array}\right.$

\item $\left\{\begin{array}{l}
\tau_1 < -\varepsilon \\
|\tau_2| \leqslant \varepsilon(1 + \alpha(-\tau_1 - \varepsilon))
\end{array}\right.$ \quad $\Rightarrow$ \quad $\left\{\begin{array}{l}
h_1 = \tau_1 + \varepsilon \\
h_2 = 0
\end{array}\right.$

\item $\left\{\begin{array}{l}
\tau_1 < -\varepsilon \\
\tau_1 < -\varepsilon(1 + \alpha(\tau_2 - \varepsilon)) \\
\tau_2 > \varepsilon(1 + \alpha(-\tau_1 - \varepsilon))
\end{array}\right.$ \quad $\Rightarrow$ \quad $\left\{\begin{array}{l}
h_1 = \tau_1 + \varepsilon(1 + \alpha(\tau_2 - \varepsilon)) \\
h_2 = \tau_2 - \varepsilon(1 + \alpha(-\tau_1 - \varepsilon))
\end{array}\right.$

\item $\left\{\begin{array}{l}
|\tau_1| \leqslant \varepsilon(1 + \alpha(\tau_2 - \varepsilon)) \\
\tau_2 > \varepsilon
\end{array}\right.$ \quad $\Rightarrow$ \quad $\left\{\begin{array}{l}
h_1 = 0 \\
h_2 = \tau_2 - \varepsilon
\end{array}\right.$

\item $\left\{\begin{array}{l}
|\tau_1| \leqslant \varepsilon \\
|\tau_2| \leqslant \varepsilon
\end{array}\right.$ \quad $\Rightarrow$ \quad $\left\{\begin{array}{l}
h_1 = 0 \\
h_2 = 0
\end{array}\right.$
\end{enumerate}}

\bigskip

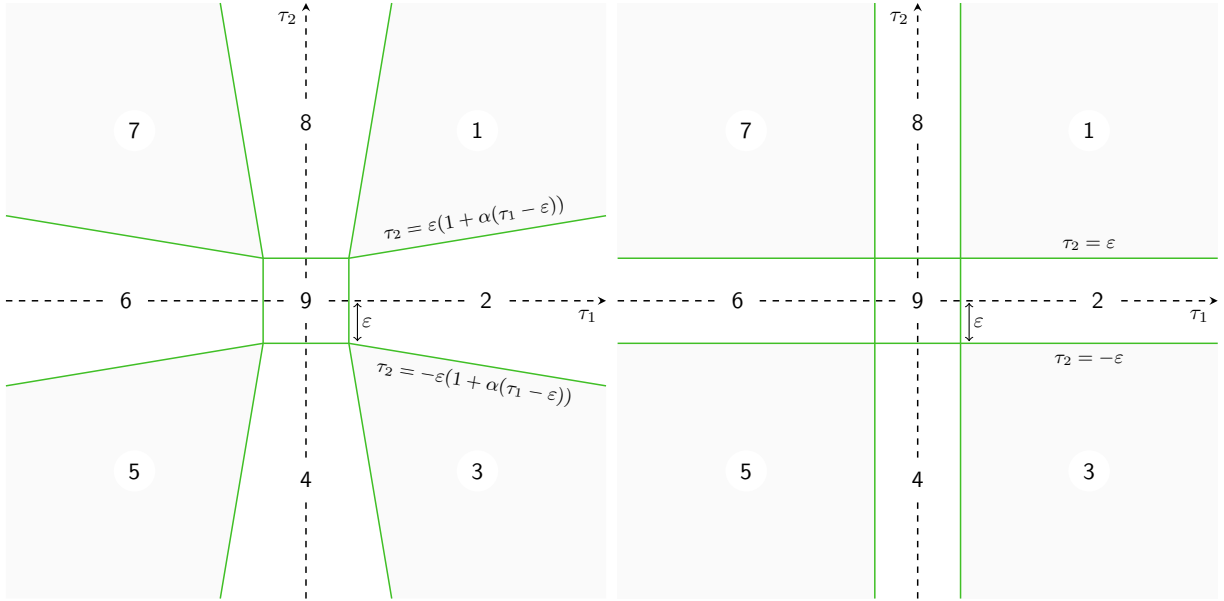
\begin{figure}[htbp]
\begin{center}
\resizebox{0.495\textwidth}{!}{\begin{tikzpicture}[scale = 0.7, line width = 1]

\tikzstyle{lignebase}=[line width = 0.7, color = vert]
\tikzstyle{texteligne}=[midway, sloped, color=black]

\fill[color = gray!4] (1,1) -- (7,2) -- (7,7) -- (2,7) -- cycle;
\fill[color = gray!4] (-1,1) -- (-2,7) -- (-7,7) -- (-7,2) -- cycle;
\fill[color = gray!4] (-1,-1) -- (-7,-2) -- (-7,-7) -- (-2,-7) -- cycle;
\fill[color = gray!4] (1,-1) -- (2,-7) -- (7,-7) -- (7,-2) -- cycle;

\draw[{->},>=stealth,dashed, line width = 0.7] (-7,0) -- (7,0);
\draw[{->},>=stealth,dashed, line width = 0.7] (0,-7) -- (0,7);
\draw[lignebase] (1,-1) -- (1,1);
\draw[lignebase] (1,1) -- (-1,1);
\draw[lignebase] (-1,1) -- (-1,-1);
\draw[lignebase] (-1,-1) -- (1,-1);

\draw[lignebase] (1,1) -- (7,2) node[texteligne, above]{\footnotesize $\tau_2 = \varepsilon(1 + \alpha(\tau_1 - \varepsilon))$};
\draw[lignebase] (1,1) -- (2,7);
\draw[lignebase] (-1,1) -- (-2,7);
\draw[lignebase] (-1,1) -- (-7,2);
\draw[lignebase] (-1,-1) -- (-7,-2);
\draw[lignebase] (-1,-1) -- (-2,-7);
\draw[lignebase] (1,-1) -- (2,-7);
\draw[lignebase] (1,-1) -- (7,-2) node[texteligne, below]{\footnotesize $\tau_2 = -\varepsilon(1 + \alpha(\tau_1 - \varepsilon))$};

\draw (7,0) node [below left] {\small $\tau_1$};
\draw (0,7) node [below left] {\small $\tau_2$};

\node[circle, fill = white] (1) at (4,4) {\textsf{1}};
\node[circle, fill = white] (2) at (4.2,0) {\textsf{2}};
\node[circle, fill = white] (3) at (4,-4) {\textsf{3}};
\node[circle, fill = white] (4) at (0,-4.2) {\textsf{4}};
\node[circle, fill = white] (5) at (-4,-4) {\textsf{5}};
\node[circle, fill = white] (6) at (-4.2,0) {\textsf{6}};
\node[circle, fill = white] (7) at (-4,4) {\textsf{7}};
\node[circle, fill = white] (8) at (0,4.2) {\textsf{8}};
\node[circle, fill = white] (9) at (0,0) {\textsf{9}};

\draw[{<->}, line width = 0.5] (1.2,-0.05) -- (1.2,-0.95);
\draw (1.1,-0.5) node [right] {\small $\varepsilon$};
\end{tikzpicture}}
\resizebox{0.495\textwidth}{!}{\begin{tikzpicture}[scale = 0.7, line width = 1]

\tikzstyle{lignebase}=[line width = 0.7, color = vert]
\tikzstyle{texteligne}=[midway, sloped, color=black]

\fill[color = gray!4] (1,1) -- (7,1) -- (7,7) -- (1,7) -- cycle;
\fill[color = gray!4] (-1,1) -- (-1,7) -- (-7,7) -- (-7,1) -- cycle;
\fill[color = gray!4] (-1,-1) -- (-7,-1) -- (-7,-7) -- (-1,-7) -- cycle;
\fill[color = gray!4] (1,-1) -- (1,-7) -- (7,-7) -- (7,-1) -- cycle;

\draw[{->},>=stealth,dashed, line width = 0.7] (-7,0) -- (7,0);
\draw[{->},>=stealth,dashed, line width = 0.7] (0,-7) -- (0,7);
\draw[lignebase] (1,-1) -- (1,1);
\draw[lignebase] (1,1) -- (-1,1);
\draw[lignebase] (-1,1) -- (-1,-1);
\draw[lignebase] (-1,-1) -- (1,-1);

\draw[lignebase] (1,1) -- (7,1) node[texteligne, above]{\footnotesize $\tau_2 = \varepsilon$};
\draw[lignebase] (1,1) -- (1,7);
\draw[lignebase] (-1,1) -- (-1,7);
\draw[lignebase] (-1,1) -- (-7,1);
\draw[lignebase] (-1,-1) -- (-7,-1);
\draw[lignebase] (-1,-1) -- (-1,-7);
\draw[lignebase] (1,-1) -- (1,-7);
\draw[lignebase] (1,-1) -- (7,-1) node[texteligne, below]{\footnotesize $\tau_2 = -\varepsilon$};

\draw (7,0) node [below left] {\small $\tau_1$};
\draw (0,7) node [below left] {\small $\tau_2$};

\node[circle, fill = white] (1) at (4,4) {\textsf{1}};
\node[circle, fill = white] (2) at (4.2,0) {\textsf{2}};
\node[circle, fill = white] (3) at (4,-4) {\textsf{3}};
\node[circle, fill = white] (4) at (0,-4.2) {\textsf{4}};
\node[circle, fill = white] (5) at (-4,-4) {\textsf{5}};
\node[circle, fill = white] (6) at (-4.2,0) {\textsf{6}};
\node[circle, fill = white] (7) at (-4,4) {\textsf{7}};
\node[circle, fill = white] (8) at (0,4.2) {\textsf{8}};
\node[circle, fill = white] (9) at (0,0) {\textsf{9}};

\draw[{<->}, line width = 0.5] (1.2,-0.05) -- (1.2,-0.95);
\draw (1.1,-0.5) node [right] {\small $\varepsilon$};
\end{tikzpicture}}
\caption{Partition of $\R^2$ associated with the proximal operator, for $\alpha>0$ (left) and $\alpha=0$ (right). Gray areas correspond to a usual gradient and white areas correspond to a threshold.}
\label{fig_prox}
\end{center}
\end{figure}

These 9 cases form a partition of $\R^2$ and are represented in Figure~\ref{fig_prox}. One can check that the case $\alpha = 0$ collapses to the usual proximal operator associated with lasso penalization.

\bigskip

To obtain the results of Fig~\ref{test_networks}, we used $\lambda = 10$, $\alpha = 5$ and $\gamma = 10^{-4}$. In a broader context, one may use standard cross-validation to derive appropriate values for $\lambda$ and $\alpha$.

\section{Dealing with real data}
\label{appendix_data}

In this section, we propose a pre-processing phase which would be required in order to apply our network inference method to real data. The first step ensures that the approximate likelihood is well-defined, while the second step consists in estimating the basal parameters appearing in functions $\koni$, $\koffi$.
Please note that inferring real networks from real data is beyond the scope of this paper and will be the subject of future papers.

\subsection{Spreading zeros}
\label{seq_spread_zeros}

The likelihood does not accept exact zeros (cf. section~\ref{seq_algo_practice}). This is not a problem with continuous-type data (for instance, based on fluorescence measurements), but it becomes one when dealing with counts (e.g. RNA-Seq). We propose to replace such zeros with relevant positive values. Recall that the PDMP focuses on the promoter and neglects the local molecular noise at the mRNA level. It is therefore natural to consider that, given a value $M>0$ of mRNA level in the PDMP, the actual number $m$ of molecules in the cell is drawn from the Poisson distribution $\Pp(M)$. Then, a possible way to replace zeros is to go backwards, i.e. to draw a value $M$ from the PDMP distribution conditioned to $m = 0$. Namely, we propose the following procedure to be applied independently for each gene:
\begin{enumerate}
\item Infer a gamma distribution $\gamma(a,b)$ (as a local approximation of the PDMP) from the whole data (possibly at a given time-point) using the standard method of moments;
\item Replace zeros with independent samples from the distribution $\gamma(a,b+1)$, conditioned to be smaller than the smallest positive value that was measured.
\end{enumerate}

This procedure ensures that zeros are replaced with very small values and that no artificial correlation is introduced. The distribution $\gamma(a,b+1)$ comes from the fact that, if $\Ll(M) = \gamma(a,b)$ and $\Ll(m | M) = \Pp(M)$, then a simple computation gives $\Ll(M|m = 0) = \gamma(a,b+1)$.

\subsection{Estimating basal parameters}

Here we describe a heuristic method to estimate the model-specific parameters (i.e. everything but the matrix $\theta$) when they cannot be measured through \emph{ad hoc} experiments, in the case of the auto-activation form~\eqref{eq_kon_full_s}-\eqref{eq_phi_full_s}. Once again we refer to section~\ref{seq_algo_practice}. Note that for the mechanistic approach to be relevant, one should know at least the ratio $d_{0,i}/d_{1,i}$, which can be obtained by measuring mRNA and protein half-lives. When even this is unavailable, we propose to use the default value $d_{0,i}/d_{1,i} = 5$ (mean value derived from the literature, cf. main text).

\bigskip

The main idea consists in noticing that, when protein $i$ is described by the auto-activation model~\eqref{eq_kon_full_s}-\eqref{eq_phi_full_s} (thus following the distribution~\eqref{eq_autoactivation_distrib}), mRNA $i$ in quasi-steady state happens to be well described by the same distribution class as~\eqref{eq_autoactivation_distrib}, with the same $m_{i,i}$ and other parameters being divided by $d_{0,i}/d_{1,i}$. More precisely, we perform the following steps:
\begin{enumerate}
\item Estimate $\widetilde{a}_{0,i} = k_{0,i}/d_{0,i}$, $\widetilde{a}_{1,i} = k_{1,i}/d_{0,i}$, $\widetilde{b}_i = \koffi/s_{0,i}$, $\widetilde{c}_i$ and $\Phi$ from the likelihood
\[f(x_i) \varpropto \sum_{r=0}^{\widetilde{c}_i} \Phi^r {x_i}^{(1- r/\widetilde{c}_i)\widetilde{a}_{0,i} + (r/\widetilde{c}_i)\widetilde{a}_{1,i} - 1} e^{-\widetilde{b}_i x_i}.\]
This can be done for instance using an EM algorithm for each value of $\widetilde{c}_i$ in some range (e.g. $\widetilde{c}_i = 1,2,\dots, 10$), and then choosing the “$\arg\max$” tuple $(\widetilde{a}_{0,i},\widetilde{a}_{1,i},\widetilde{b}_i,\widetilde{c}_i,\Phi)$. Afterwards, $\widetilde{a}_{0,i}$, $\widetilde{a}_{1,i}$, $\widetilde{b}_i$ and $\widetilde{c}_i$ are stored ($\Phi$ only serves this step).

\item Consistently with the definition of the model, we set $m_{i,i} = (\widetilde{a}_{1,i}-\widetilde{a}_{0,i})/\widetilde{c}_i$,
\[a_{0,i} = \frac{k_{0,i}}{d_{1,i}} = \frac{d_{0,i}}{d_{1,i}} \cdot \widetilde{a}_{0,i}, \quad a_{1,i} = \frac{k_{1,i}}{d_{1,i}} = \frac{d_{0,i}}{d_{1,i}} \cdot \widetilde{a}_{1,i}, \quad c_i = \frac{k_{1,i} - k_{0,i}}{d_{1,i}m_{i,i}} = \frac{d_{0,i}}{d_{1,i}} \cdot \widetilde{c}_i\]
and we choose $b_i = \frac{d_{0,i}}{d_{1,i}}\cdot\widetilde{b}_i$. Then we define, as an approximation of~\eqref{eq_s_sym} in the bursty regime,
\[s_{i,i} = \frac{1}{b_i}\left(\frac{\Gamma(a_{1,i})}{\Gamma(a_{0,i})}\right)^{1/(a_{1,i} - a_{0,i})} .\]
Note that such $b_i$ is not the “true” value regarding section~\ref{seq_likelihood_autoactiv_gamma}, as we would need to know $\frac{d_{1,i}}{s_{1,i}}$ to apply the formula $b_i = \frac{d_{1,i}}{s_{1,i}} \cdot \frac{d_{0,i}}{d_{1,i}}\cdot\widetilde{b}_i$. Fortunately, the network inference does not depend on this scale parameter since the Hill threshold $s_{i,i}$ is proportional to $1/b_i$.

\item Last step consists in extrapolating $m_{i,i}$ and $s_{i,i}$ to the remaining unknown parameters $m_{i,j}$ and $s_{i,j}$ (describing how gene $j$ influences gene $i$). Since the crucial point is their coherence with respect to the range of protein $j$, a relevant choice without additional knowledge is, for all $i \neq j$,
\[m_{i,j} = m_{j,j} \quad \text{and} \quad s_{i,j} = s_{j,j} . \]
\end{enumerate}

\section{Parameter values}
\label{appendix_parameters}

\subsection{Models}


\begin{table}[ht]
\caption{General parameters used in the examples. The $s_{i,j}$ correspond to the normalized model: counterparts in absolute protein numbers are $\overline{s}_{i,j} = s_{i,j}\times (s_0s_1)/(d_0d_1) = \num{2e3}$ for $i\neq j$ and $\overline{s}_{i,i} = s_{i,i}\times (s_0s_1)/(d_0d_1) = \num{1.9e4}$.}
\begin{center}
\begin{tabular}{lll}
\hline
Parameter & Value & Units \\
\hline
$s_0$ & $10^3$ & \si{mRNA\cdot h^{-1}} \\
$s_1$ & $10$ & \si{protein\cdot h^{-1} \cdot mRNA^{-1}} \\
$d_0$ & $0.5$ & \si{h^{-1}} \\
$d_1$ & $0.1$ & \si{h^{-1}} \\
$k_0$ & $0.34$ & \si{h^{-1}} \\
$k_1$ & $2.15$ & \si{h^{-1}} \\
$\koff$ & $10$ & \si{h^{-1}} \\
$m_{i,j}$ & $2$ for $i\neq j$ & -- \\
$m_{i,i}$ & $2$ (Fig 5, 6) or $3$ (Fig 8) & -- \\
$s_{i,j}$ & $0.01$ for $i\neq j$ & proteins (normalized) \\
$s_{i,i}$ & $0.095$ from eq. \eqref{eq_s_sym} & proteins (normalized) \\
\hline
\end{tabular}
\label{tab_parameters_base}
\end{center}
\end{table}

\clearpage

\begin{table}[ht]
\caption{Network parameters used in the examples.}
\begin{center}
\begin{tabular}{ccccc}
\hline
Fig 5, 6 & $\theta_{1,1}$ & $\theta_{1,2}$ & $\theta_{2,1}$ & $\theta_{2,2}$ \\
\hline
& $4$ & $-8$ & $-8$ & $4$ \\
\hline
Fig 8 & $\theta_{1,1}$ & $\theta_{1,2}$ & $\theta_{2,1}$ & $\theta_{2,2}$ \\
\hline
1 & $0$ & $0$ & $0$ & $0$ \\
2 & $0$ & $0$ & $1$ & $0$ \\
3 & $0$ & $1$ & $0$ & $0$ \\
4 & $-0.1$ & $1$ & $1$ & $-0.1$ \\
5 & $0$ & $0$ & $-1$ & $0$ \\
6 & $0$ & $-1$ & $0$ & $0$ \\
7 & $0$ & $-1$ & $-1$ & $0$ \\
\hline
\end{tabular}
\label{tab_parameters_networks}
\end{center}
\end{table}


\subsection{Results}

\begin{table}[hb]
\small
\caption{Inferred network parameters used to generate Fig 8b. Each row refers to one of the ten datasets generated for testing. Colors indicate whether the parameters represent the correct topology (blue) or another one (orange) regarding the true networks (cf. Table~\ref{tab_parameters_networks}).}
\begin{center}
\begin{tabular}{|r|cc|}
\hline \multicolumn{3}{|c|}{Network 1} \\
\hline
& $\theta_{1,2}$ & $\theta_{2,1}$ \\
\hline
1 & {\color{bleu}$0$} & {\color{bleu}$0$} \\
2 & {\color{bleu}$0$} & {\color{bleu}$0$} \\
3 & {\color{bleu}$0$} & {\color{bleu}$0$} \\
4 & {\color{bleu}$0$} & {\color{bleu}$0$} \\
5 & {\color{orange}$0.10$} & {\color{orange}$0$} \\
6 & {\color{bleu}$0$} & {\color{bleu}$0$} \\
7 & {\color{bleu}$0$} & {\color{bleu}$0$} \\
8 & {\color{bleu}$0$} & {\color{bleu}$0$} \\
9 & {\color{bleu}$0$} & {\color{bleu}$0$} \\
10 & {\color{bleu}$0$} & {\color{bleu}$0$} \\
\hline
\end{tabular}

\begin{tabular}{|r|cc|}
\hline \multicolumn{3}{|c|}{Network 2} \\
\hline
& $\theta_{1,2}$ & $\theta_{2,1}$ \\
\hline
1 & {\color{bleu}$0$} & {\color{bleu}$0.13$} \\
2 & {\color{bleu}$0$} & {\color{bleu}$0.18$} \\
3 & {\color{bleu}$0$} & {\color{bleu}$0.14$} \\
4 & {\color{bleu}$0$} & {\color{bleu}$0.17$} \\
5 & {\color{orange}$0$} & {\color{orange}$0$} \\
6 & {\color{bleu}$0$} & {\color{bleu}$0.20$} \\
7 & {\color{bleu}$0$} & {\color{bleu}$0.18$} \\
8 & {\color{bleu}$0$} & {\color{bleu}$0.11$} \\
9 & {\color{bleu}$0$} & {\color{bleu}$0.15$} \\
10 & {\color{orange}$0$} & {\color{orange}$0$} \\
\hline
\end{tabular}

\begin{tabular}{|r|cc|}
\hline \multicolumn{3}{|c|}{Network 3} \\
\hline
& $\theta_{1,2}$ & $\theta_{2,1}$ \\
\hline
1 & {\color{orange}$0$} & {\color{orange}$0$} \\
2 & {\color{bleu}$0.12$} & {\color{bleu}$0$} \\
3 & {\color{orange}$0.26$} & {\color{orange}$0.10$} \\
4 & {\color{bleu}$0.19$} & {\color{bleu}$0$} \\
5 & {\color{bleu}$0.21$} & {\color{bleu}$0$} \\
6 & {\color{orange}$0.36$} & {\color{orange}$0.15$} \\
7 & {\color{bleu}$0.11$} & {\color{bleu}$0$} \\
8 & {\color{bleu}$0.11$} & {\color{bleu}$0$} \\
9 & {\color{bleu}$0.19$} & {\color{bleu}$0$} \\
10 & {\color{bleu}$0.12$} & {\color{bleu}$0$} \\
\hline
\end{tabular}

\begin{tabular}{|r|cc|}
\hline \multicolumn{3}{|c|}{Network 4} \\
\hline
& $\theta_{1,2}$ & $\theta_{2,1}$ \\
\hline
1 & {\color{bleu}$0.28$} & {\color{bleu}$0.22$} \\
2 & {\color{bleu}$0.25$} & {\color{bleu}$0.15$} \\
3 & {\color{bleu}$0.23$} & {\color{bleu}$0.21$} \\
4 & {\color{bleu}$0.20$} & {\color{bleu}$0.15$} \\
5 & {\color{bleu}$0.18$} & {\color{bleu}$0.23$} \\
6 & {\color{bleu}$0.23$} & {\color{bleu}$0.27$} \\
7 & {\color{bleu}$0.17$} & {\color{bleu}$0.16$} \\
8 & {\color{bleu}$0.14$} & {\color{bleu}$0.13$} \\
9 & {\color{bleu}$0.29$} & {\color{bleu}$0.21$} \\
10 & {\color{bleu}$0.24$} & {\color{bleu}$0.21$} \\
\hline
\end{tabular}
 \\ \vspace{2mm}
\begin{tabular}{|r|cc|}
\hline \multicolumn{3}{|c|}{Network 5} \\
\hline
& $\theta_{1,2}$ & $\theta_{2,1}$ \\
\hline
1 & {\color{orange}$0$} & {\color{orange}$0$} \\
2 & {\color{bleu}$0$} & {\color{bleu}$-0.12$} \\
3 & {\color{bleu}$0$} & {\color{bleu}$-0.10$} \\
4 & {\color{bleu}$0$} & {\color{bleu}$-0.16$} \\
5 & {\color{orange}$-0.10$} & {\color{orange}$-0.21$} \\
6 & {\color{bleu}$0$} & {\color{bleu}$-0.18$} \\
7 & {\color{bleu}$0$} & {\color{bleu}$-0.13$} \\
8 & {\color{bleu}$0$} & {\color{bleu}$-0.09$} \\
9 & {\color{orange}$0$} & {\color{orange}$0$} \\
10 & {\color{bleu}$0$} & {\color{bleu}$-0.18$} \\
\hline
\end{tabular}

\begin{tabular}{|r|cc|}
\hline \multicolumn{3}{|c|}{Network 6} \\
\hline
& $\theta_{1,2}$ & $\theta_{2,1}$ \\
\hline
1 & {\color{bleu}$-0.14$} & {\color{bleu}$0$} \\
2 & {\color{bleu}$-0.11$} & {\color{bleu}$0$} \\
3 & {\color{bleu}$-0.13$} & {\color{bleu}$0$} \\
4 & {\color{bleu}$-0.15$} & {\color{bleu}$0$} \\
5 & {\color{bleu}$-0.21$} & {\color{bleu}$0$} \\
6 & {\color{bleu}$-0.16$} & {\color{bleu}$0$} \\
7 & {\color{orange}$-0.29$} & {\color{orange}$-0.09$} \\
8 & {\color{orange}$0$} & {\color{orange}$0$} \\
9 & {\color{bleu}$-0.13$} & {\color{bleu}$0$} \\
10 & {\color{bleu}$-0.21$} & {\color{bleu}$0$} \\
\hline
\end{tabular}

\begin{tabular}{|r|cc|}
\hline \multicolumn{3}{|c|}{Network 7} \\
\hline
& $\theta_{1,2}$ & $\theta_{2,1}$ \\
\hline
1 & {\color{bleu}$-0.20$} & {\color{bleu}$-0.24$} \\
2 & {\color{bleu}$-0.34$} & {\color{bleu}$-0.34$} \\
3 & {\color{bleu}$-0.17$} & {\color{bleu}$-0.14$} \\
4 & {\color{bleu}$-0.23$} & {\color{bleu}$-0.20$} \\
5 & {\color{bleu}$-0.17$} & {\color{bleu}$-0.18$} \\
6 & {\color{bleu}$-0.19$} & {\color{bleu}$-0.16$} \\
7 & {\color{bleu}$-0.21$} & {\color{bleu}$-0.21$} \\
8 & {\color{bleu}$-0.30$} & {\color{bleu}$-0.31$} \\
9 & {\color{bleu}$-0.11$} & {\color{bleu}$-0.13$} \\
10 & {\color{bleu}$-0.11$} & {\color{bleu}$-0.17$} \\
\hline
\end{tabular}

\vspace{-27mm}
\end{center}
\label{tab_inferred_networks}
\end{table}

\clearpage

\begin{table}[htb]
\small
\caption{Example of inferred networks in the presence of dropouts (30\% of the whole dataset) generated by applying a Poisson noise and then a threshold to the “perfect” data. Such zeros where replaced using the procedure described in section~\ref{seq_spread_zeros} before inferring the networks.}
\begin{center}
\begin{tabular}{|r|cc|}
\hline \multicolumn{3}{|c|}{Network 1} \\
\hline
& $\theta_{1,2}$ & $\theta_{2,1}$ \\
\hline
1 & {\color{bleu}$0$} & {\color{bleu}$0$} \\
2 & {\color{orange}$0$} & {\color{orange}$0.10$} \\
3 & {\color{bleu}$0$} & {\color{bleu}$0$} \\
4 & {\color{bleu}$0$} & {\color{bleu}$0$} \\
5 & {\color{orange}$0.10$} & {\color{orange}$0$} \\
6 & {\color{bleu}$0$} & {\color{bleu}$0$} \\
7 & {\color{orange}$0.09$} & {\color{orange}$0.16$} \\
8 & {\color{bleu}$0$} & {\color{bleu}$0$} \\
9 & {\color{bleu}$0$} & {\color{bleu}$0$} \\
10 & {\color{bleu}$0$} & {\color{bleu}$0$} \\
\hline
\end{tabular}

\begin{tabular}{|r|cc|}
\hline \multicolumn{3}{|c|}{Network 2} \\
\hline
& $\theta_{1,2}$ & $\theta_{2,1}$ \\
\hline
1 & {\color{bleu}$0$} & {\color{bleu}$0.27$} \\
2 & {\color{bleu}$0$} & {\color{bleu}$0.16$} \\
3 & {\color{bleu}$0$} & {\color{bleu}$0.31$} \\
4 & {\color{bleu}$0$} & {\color{bleu}$0.44$} \\
5 & {\color{bleu}$0$} & {\color{bleu}$0.27$} \\
6 & {\color{bleu}$0$} & {\color{bleu}$0.55$} \\
7 & {\color{bleu}$0$} & {\color{bleu}$0.23$} \\
8 & {\color{bleu}$0$} & {\color{bleu}$0.21$} \\
9 & {\color{bleu}$0$} & {\color{bleu}$0.36$} \\
10 & {\color{orange}$0$} & {\color{orange}$0$} \\
\hline
\end{tabular}

\begin{tabular}{|r|cc|}
\hline \multicolumn{3}{|c|}{Network 3} \\
\hline
& $\theta_{1,2}$ & $\theta_{2,1}$ \\
\hline
1 & {\color{bleu}$0.31$} & {\color{bleu}$0$} \\
2 & {\color{bleu}$0.23$} & {\color{bleu}$0$} \\
3 & {\color{orange}$0.32$} & {\color{orange}$0.12$} \\
4 & {\color{bleu}$0.30$} & {\color{bleu}$0$} \\
5 & {\color{orange}$0.39$} & {\color{orange}$0.11$} \\
6 & {\color{orange}$0.40$} & {\color{orange}$0.16$} \\
7 & {\color{orange}$0.19$} & {\color{orange}$0.11$} \\
8 & {\color{bleu}$0.14$} & {\color{bleu}$0$} \\
9 & {\color{orange}$0.35$} & {\color{orange}$0.16$} \\
10 & {\color{bleu}$0.19$} & {\color{bleu}$0$} \\
\hline
\end{tabular}

\begin{tabular}{|r|cc|}
\hline \multicolumn{3}{|c|}{Network 4} \\
\hline
& $\theta_{1,2}$ & $\theta_{2,1}$ \\
\hline
1 & {\color{bleu}$0.43$} & {\color{bleu}$0.41$} \\
2 & {\color{bleu}$0.44$} & {\color{bleu}$0.26$} \\
3 & {\color{bleu}$0.36$} & {\color{bleu}$0.28$} \\
4 & {\color{bleu}$0.44$} & {\color{bleu}$0.23$} \\
5 & {\color{bleu}$0.38$} & {\color{bleu}$0.51$} \\
6 & {\color{bleu}$0.42$} & {\color{bleu}$0.48$} \\
7 & {\color{bleu}$0.48$} & {\color{bleu}$0.34$} \\
8 & {\color{bleu}$0.36$} & {\color{bleu}$0.32$} \\
9 & {\color{bleu}$0.59$} & {\color{bleu}$0.43$} \\
10 & {\color{bleu}$0.41$} & {\color{bleu}$0.34$} \\
\hline
\end{tabular}
 \\ \vspace{2mm}
\begin{tabular}{|r|cc|}
\hline \multicolumn{3}{|c|}{Network 5} \\
\hline
& $\theta_{1,2}$ & $\theta_{2,1}$ \\
\hline
1 & {\color{orange}$0$} & {\color{orange}$0$} \\
2 & {\color{bleu}$0$} & {\color{bleu}$-0.26$} \\
3 & {\color{orange}$-0.09$} & {\color{orange}$-0.21$} \\
4 & {\color{orange}$-0.11$} & {\color{orange}$-0.35$} \\
5 & {\color{orange}$-0.30$} & {\color{orange}$-0.47$} \\
6 & {\color{bleu}$0$} & {\color{bleu}$-0.32$} \\
7 & {\color{bleu}$0$} & {\color{bleu}$-0.22$} \\
8 & {\color{bleu}$0$} & {\color{bleu}$-0.19$} \\
9 & {\color{bleu}$0$} & {\color{bleu}$-0.27$} \\
10 & {\color{bleu}$0$} & {\color{bleu}$-0.38$} \\
\hline
\end{tabular}

\begin{tabular}{|r|cc|}
\hline \multicolumn{3}{|c|}{Network 6} \\
\hline
& $\theta_{1,2}$ & $\theta_{2,1}$ \\
\hline
1 & {\color{bleu}$-0.18$} & {\color{bleu}$0$} \\
2 & {\color{bleu}$-0.25$} & {\color{bleu}$0$} \\
3 & {\color{bleu}$-0.28$} & {\color{bleu}$0$} \\
4 & {\color{bleu}$-0.25$} & {\color{bleu}$0$} \\
5 & {\color{bleu}$-0.34$} & {\color{bleu}$0$} \\
6 & {\color{bleu}$-0.26$} & {\color{bleu}$0$} \\
7 & {\color{orange}$-0.47$} & {\color{orange}$-0.18$} \\
8 & {\color{bleu}$-0.09$} & {\color{bleu}$0$} \\
9 & {\color{bleu}$-0.22$} & {\color{bleu}$0$} \\
10 & {\color{orange}$-0.38$} & {\color{orange}$-0.12$} \\
\hline
\end{tabular}

\begin{tabular}{|r|cc|}
\hline \multicolumn{3}{|c|}{Network 7} \\
\hline
& $\theta_{1,2}$ & $\theta_{2,1}$ \\
\hline
1 & {\color{bleu}$-0.22$} & {\color{bleu}$-0.24$} \\
2 & {\color{bleu}$-0.5$} & {\color{bleu}$-0.50$} \\
3 & {\color{bleu}$-0.25$} & {\color{bleu}$-0.21$} \\
4 & {\color{bleu}$-0.26$} & {\color{bleu}$-0.22$} \\
5 & {\color{bleu}$-0.12$} & {\color{bleu}$-0.14$} \\
6 & {\color{bleu}$-0.20$} & {\color{bleu}$-0.18$} \\
7 & {\color{bleu}$-0.30$} & {\color{bleu}$-0.29$} \\
8 & {\color{bleu}$-0.44$} & {\color{bleu}$-0.45$} \\
9 & {\color{bleu}$-0.15$} & {\color{bleu}$-0.18$} \\
10 & {\color{bleu}$-0.11$} & {\color{bleu}$-0.17$} \\
\hline
\end{tabular}

\end{center}
\label{tab_inferred_networks_dropout}
\end{table}



\end{appendices}

\clearpage

\renewcommand{\baselinestretch}{1}

{\small


}

\end{document}